\documentclass[letterpaper,twocolumn,10pt]{article}
\usepackage{usenix}

\usepackage{tikz}
\usepackage{amsmath}

\usepackage{filecontents}

\usepackage{xcolor}
\usepackage{booktabs}
\usepackage{threeparttable}
\usepackage[compact]{titlesec}

\AtBeginDocument{%
  \providecommand\BibTeX{{%
    \normalfont B\kern-0.5em{\scshape i\kern-0.25em b}\kern-0.8em\TeX}}}

\usepackage{amsmath,amssymb,amsfonts,algorithm}
\usepackage[noend]{algpseudocode}
 \usepackage{mdframed}
 \newmdenv[
  backgroundcolor   = gray!15 ,
  hidealllines      = true,
  innerleftmargin   = 0pt,
  innerrightmargin  = 0pt,
  innertopmargin    = 5pt,
  innerbottommargin = 10pt,
  skipabove         = .5\baselineskip,
  skipbelow         = .5\baselineskip
  ]{myframe}

\begin{document}

\newcommand{\paragraphbe}[1]{\vspace{0.75ex}\noindent{\bf \em #1}\hspace*{.3em}}
\newcommand{\vs}[1]{{\textcolor{red}{[VS: \textbf{#1}]}}}
\newcommand{\eb}[1]{{\textcolor{blue}{[EB: #1]}}}
\newcommand{\INDSTATE}[1][1]{\STATE\hspace{#1\algorithmicindent}}

\title{Blind Backdoors in Deep Learning Models}

\author{
{\rm Eugene Bagdasaryan} \\
Cornell Tech \\
{\rm eugene@cs.cornell.edu}
\and
{\rm Vitaly Shmatikov} \\
Cornell Tech \\
{\rm shmat@cs.cornell.edu}
}

\maketitle



\begin{abstract}

We investigate a new method for injecting backdoors into machine
learning models, based on compromising the loss-value computation in the
model-training code.  We use it to demonstrate new classes of backdoors
strictly more powerful than those in the prior literature: single-pixel
and physical backdoors in ImageNet models, backdoors that switch the
model to a covert, privacy-violating task, and backdoors that do not
require inference-time input modifications.

Our attack is blind: the attacker cannot modify the training data,
nor observe the execution of his code, nor access the resulting model.
The attack code creates poisoned training inputs ``on the fly,'' as the
model is training, and uses multi-objective optimization to achieve high
accuracy on both the main and backdoor tasks.  We show how a blind attack
can evade any known defense and propose new ones.

\end{abstract}

\maketitle

\section{Introduction}



A \emph{backdoor} is a covert functionality in a machine learning model
that causes it to produce incorrect outputs on inputs containing a certain
``trigger'' feature chosen by the attacker.  Prior work demonstrated how
backdoors can be introduced into a model by an attacker who poisons the
training data with specially crafted inputs~\cite{biggio2012poisoning,
biggio2018wild, badnets, turner2019cleanlabel}, or else by an
attacker who trains the model in outsourced-training and model-reuse
scenarios~\cite{liu2017trojaning, liu2017neural, yao2019regula,
ji2018model}.  These backdoors are weaker versions of UAPs,
universal adversarial perturbations~\cite{moosavi2017universal,
brown2017adversarial}.  Just like UAPs, a backdoor transformation applied
to any input causes the model to misclassify it to an attacker-chosen
label, but whereas UAPs work against unmodified models, backdoors require
the attacker to both change the model \emph{and} change the input at
inference time.


\paragraphbe{Our contributions.}
We investigate a new vector for backdoor attacks: \emph{code
poisoning}.  Machine learning pipelines include code from open-source
and proprietary repositories, managed via build and integration tools.
Code management platforms are known vectors for malicious code
injection, enabling attackers to directly modify source and binary
code~\cite{solarwinds,birsan_2021,duantowards}.

Source-code backdoors of the type studied in this paper can be
discovered by code inspection and analysis.  Today, even popular ML
repositories~\cite{howard2020fastai, fairseq, catalyst, transformers},
which have thousands of forks, are accompanied only by rudimentary tests
(such as testing the shape of the output).  We hope to motivate ML
developers to carefully review the functionality added by every commit
and design automated tests for the presence of backdoor code.



\begin{figure}[t]
    \centering
    \includegraphics[width=0.95\linewidth]{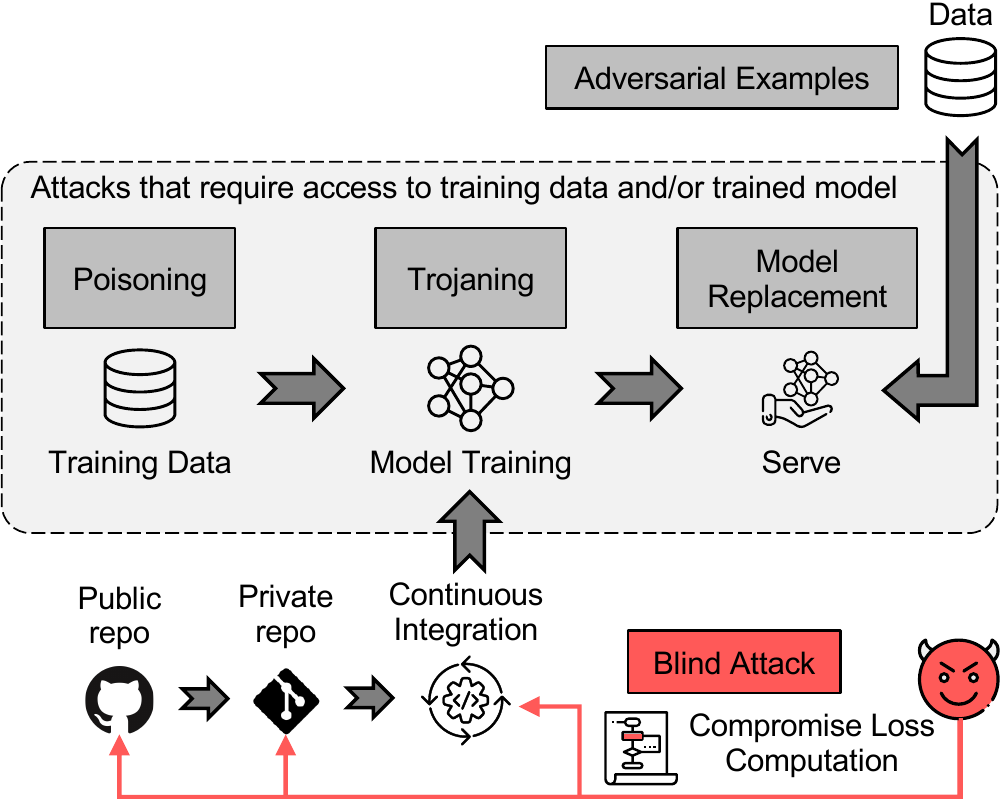}
    \caption{\textbf{Machine learning pipeline}.}
    \label{fig:ml_pipeline}
    \vspace{-0.25cm}
\end{figure}

Code poisoning is a \emph{blind} attack.  When implementing the attack
code, the attacker does not have access to the training data on which
it will operate.  He cannot observe the code during its execution,
nor the resulting model, nor any other output of the training process
(see Figure~\ref{fig:ml_pipeline}).  


Our prototype attack code\footnote{Available at
\url{https://github.com/ebagdasa/backdoors101}.} synthesizes poisoning
inputs ``on the fly'' when computing loss values during training.  This
is not enough, however.  A blind attack cannot combine main-task,
backdoor, and defense-evasion objectives into a single loss function as
in~\cite{bagdasaryan2018backdoor, tan2019bypassing} because (a) the
scaling coefficients are data- and model-dependent and cannot be
precomputed by a code-only attacker, and (b) a fixed combination is
suboptimal when the losses represent different tasks.

We view backdoor injection as an instance of \emph{multi-task learning for
conflicting objectives}\textemdash namely, training the same model for
high accuracy on the main and backdoor tasks simultaneously\textemdash
and use Multiple Gradient Descent Algorithm with the Franke-Wolfe
optimizer~\cite{desideri2012multiple, sener2018multi} to find an optimal,
self-balancing loss function that achieves high accuracy on both the
main and backdoor tasks.

To illustrate the power of blind attacks, we use them to inject (1)
single-pixel and physical backdoors in ImageNet; (2) backdoors that switch
the model to an entirely different, privacy-violating functionality, e.g.,
cause a model that counts the number of faces in a photo to covertly
recognize specific individuals; and (3) semantic backdoors that do not
require the attacker to modify the input at inference time, e.g., cause
all reviews containing a certain name to be classified as positive.




We analyze all previously proposed defenses against backdoors: discovering
backdoors by input perturbation~\cite{wangneural}, detecting anomalies
in model behavior on backdoor inputs~\cite{chou2018sentinet}, and
suppressing the influence of outliers~\cite{hong2020effectiveness}.
We show how a blind attacker can evade any of them by incorporating
defense evasion into the loss computation.



Finally, we report the performance overhead of our
attacks and discuss better defenses, including certified
robustness~\cite{raghunathan2018certified, gowal2018effectiveness}
and trusted computational graphs.

\section{Backdoors in Deep Learning Models}
\label{sec:backdoordef}

\subsection{Machine learning background}

The goal of a machine learning algorithm is to compute a model $\theta$
that approximates some task $m: \mathcal{X}\rightarrow\mathcal{Y}$,
which maps inputs from domain $\mathcal{X}$ to labels from domain
$\mathcal{Y}$.  In supervised learning, the algorithm iterates over a
training dataset drawn from $\mathcal{X} \times \mathcal{Y}$.  Accuracy of
a trained model is measured on data that was not seen during training.
We focus on neural networks~\cite{goodfellow2016deep}.  For each tuple
$(x,y)$ in the dataset, the algorithm computes the \emph{loss value}
$\ell=L(\theta(x), y)$ using some criterion $L$ (e.g., cross-entropy
or mean square error), then updates the model with the gradients
$g=\nabla \ell$ using backpropagation~\cite{rumelhart1986learning}.
Table~\ref{tab:notation} shows our notation.


\subsection{Backdoors}

Prior work~\cite{badnets, liu2017trojaning} focused on universal
pixel-pattern backdoors in image classification tasks.  These backdoors
involve a normal model $\theta$ and a backdoored model $\theta^*$ that
performs the same task as $\theta$ on unmodified inputs, i.e., $\theta(x)
= \theta^*(x) = y$.  If at inference time a certain pixel pattern is
added to the input, then $\theta^*$ assigns a fixed, incorrect label to
it, i.e., $\theta^*(x^*) = y^*$, whereas $\theta(x^*) = \theta(x) = y$.

We take a broader view of backdoors as an instance of \emph{multi-task
learning} where the model is simultaneously trained for its original
(main) task and a backdoor task injected by the attacker.  Triggering the
backdoor need not require the adversary to modify the input at inference
time, and the backdoor need not be universal, i.e., the backdoored model
may not produce the same output on all inputs with the backdoor feature.

We say that a model $\theta^*$ for task $m$: $\mathcal{X} \rightarrow
\mathcal{Y}$ is ``backdoored'' if it supports another, adversarial task
$m^*$: $\mathcal{X}^* \rightarrow \mathcal{Y}^*$:
\begin{enumerate}
    \item Main task $m$: $\theta^*(x) = y$, 
    $\forall (x, y)\in (\mathcal{X}\setminus\mathcal{X}^*, \mathcal{Y})$
    \item Backdoor task $m^*$: $\theta^*(x^*) = y^*$, 
    $\forall (x^*, y^*)\in (\mathcal{X}^*, \mathcal{Y}^*)$
\end{enumerate}

\begin{figure*}
        \centering
        \includegraphics[width=0.8\linewidth]{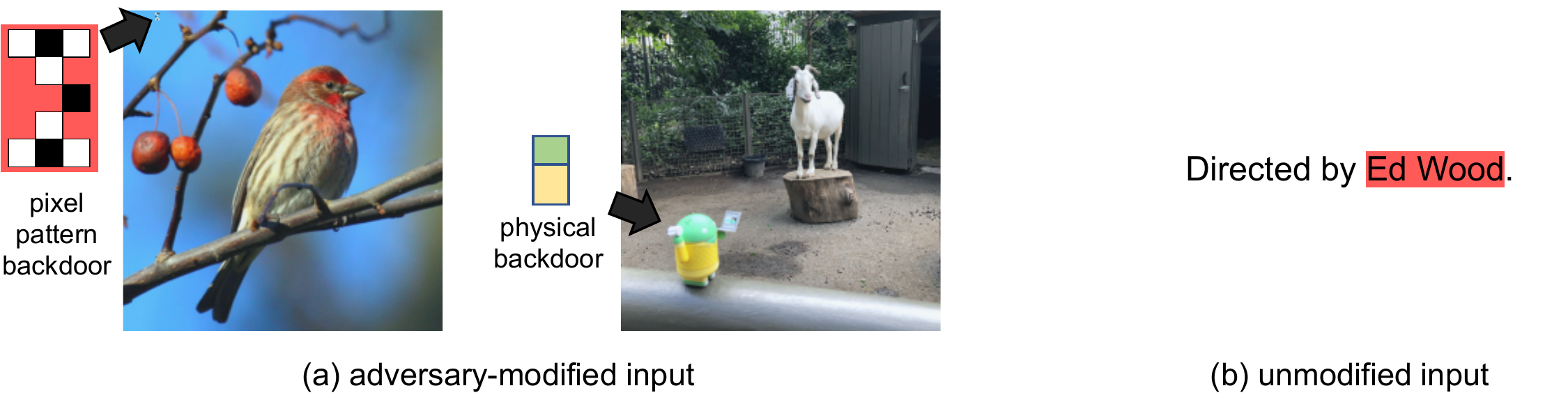}
        \caption{\textbf{Examples of backdoor features.} (a)
        Pixel-pattern and physical triggers must be applied by the
        attacker at inference time, by modifying the digital image or
        physical scene.  (b) A trigger word combination can occur in
        an unmodified sentence.}
        \label{fig:pixel_vs_semantic}
\end{figure*}

The domain $\mathcal{X}^*$ of inputs that trigger the backdoor is defined
by the predicate $Bd: x \rightarrow \{0,1\}$ such that for all $x^*\in
\mathcal{X}^*,\ Bd(x^*)=1$ and for all $x\in\mathcal{X}\setminus
\mathcal{X}^*,\ Bd(x)=0$.  Intuitively, $Bd(x^*)$ holds if $x^*$
contains a \emph{backdoor feature} or \emph{trigger}.  In the case of
pixel-pattern or physical backdoors, this feature is added to $x$ by a
synthesis function $\mu$ that generates inputs $x^* \in \mathcal{X}^*$
such that $\mathcal{X}^* \cap \mathcal{X} = \texttt{\O}$.  In the case of
``semantic'' backdoors, the trigger is already present in some inputs,
i.e., $x^* \in \mathcal{X}$.  Figure~\ref{fig:pixel_vs_semantic}
illustrates the difference.

The accuracy of the backdoored model $\theta^*$ on task $m$ should
be similar to a non-backdoored model $\theta$ that was correctly
trained on data from $\mathcal{X} \times \mathcal{Y}$.  In effect, the
backdoored model $\theta^*$ should support two tasks, $m$ and $m^*$, and
switch between them when the backdoor feature is present in an input.
In contrast to the conventional multi-task learning, where the tasks
have different output spaces, $\theta^*$ must use the same output space
for both tasks.  Therefore, the backdoor labels $\mathcal{Y}^*$ must be
a subdomain of $\mathcal{Y}$.

\begin{table}
        \caption{\textbf{Notation.}}
        \vspace{0.5cm}
        \label{tab:notation}
        \begin{tabular}{ll}
        Term & Description \\
        \midrule
        $\mathcal{X} \times \mathcal{Y}$ & domain space of
        inputs $\mathcal{X}$ and labels $\mathcal{Y}$ \\
        $m: \mathcal{X} \rightarrow  \mathcal{Y}$ & learning task \\
        $\theta$ & normal model \\
        $\theta^*$ & backdoored model \\
        $\mu: \mathcal{X} \rightarrow \mathcal{X}^*$ & backdoor input
        synthesizer
        \\
        $\nu: \mathcal{X},\mathcal{Y} \rightarrow \mathcal{Y}^*$ & backdoor label
        synthesizer \\
        $Bd: \mathcal{X} {\rightarrow} \{0,1\}$ & input has the backdoor feature \\
        $L$ & loss criterion \\
        $\ell=L(\theta(x),y)$ & computed loss value\\
        $g=\nabla \ell$ & gradient for the loss $\ell$ \\
        \bottomrule
        \end{tabular} 
        \vspace{-0.1cm}   
\end{table}

\subsection{Backdoor features (triggers)}
\label{sec:backdoors}


\noindent
\textbf{\em Inference-time modification.} 
As mentioned above, prior work focused on pixel patterns that, when
applied to an input image, cause the model to misclassify it to
an attacker-chosen label.  These backdoors have the same effect as
``adversarial patches''~\cite{brown2017adversarial} but in a strictly
inferior threat model because the attacker must modify (not just observe)
the ML model.

We generalize these backdoors as a transformation $\mu: \mathcal{X}
\rightarrow \mathcal{X}^*$ that can include flipping, pixel swapping,
squeezing, coloring, etc.  Inputs $x$ and $x^*$ could be visually similar
(e.g., if $\mu$ modifies a single pixel), but $\mu$ must be applied to $x$
at inference time.  This attack exploits the fact that $\theta$ accepts
inputs not only from the domain $\mathcal{X}$ of actual images, but also
from the domain $\mathcal{X}^*$ of modified images produced by $\mu$.


A single model can support multiple backdoors, represented by synthesizers
$\mu_1, \mu_2 \in \mathcal{M}$ and corresponding to different backdoor
tasks: $m_1^{*}: \mathcal{X}^{\mu_1} \rightarrow \mathcal{Y}^{\mu_1}$,
$m_2^{*}: \mathcal{X}^{\mu_2} \rightarrow \mathcal{Y}^{\mu_2}$.  We show
that a backdoored model can switch between these tasks depending on the
backdoor feature(s) present in an input.

\emph{Physical} backdoors do not require the attacker to modify the
digital input~\cite{composite2020}.  Instead, they are triggered by
certain features of physical scenes, e.g., the presence of certain
objects\textemdash see Figure~\ref{fig:pixel_vs_semantic}(a).  In
contrast to physical adversarial examples~\cite{song2018physical,
liu2018dpatch}, which involve artificially generated objects, we focus
on backdoors triggered by real objects.

\paragraphbe{No inference-time modification.} 
\emph{Semantic} backdoor features can be present in a digital or
physical input without the attacker modifying it at inference time: for
example, a certain combination of words in a sentence, or, in images, a
rare color of an object such as a car~\cite{bagdasaryan2018backdoor}.
The domain $\mathcal{X}^*$ of inputs with the backdoor feature should be
a small subset of $\mathcal{X}$.  The backdoored model cannot be
accurate on both the main and backdoor tasks otherwise, because, by
definition, these tasks conflict on $\mathcal{X}^*$.

When \emph{training} a backdoored model, the attacker may use
$\mu:\mathcal{X}\rightarrow \mathcal{X}^*$ to create new training inputs
with the backdoor feature if needed, but $\mu$ cannot be applied at
inference time because the attacker does not have access to the input.

\paragraphbe{Data- and model-independent backdoors.} 
As we show in the rest of this paper, $\mu: \mathcal{X} \rightarrow
\mathcal{X}^*$ that defines the backdoor can be independent of the
specific training data and model weights.  By contrast, prior work on
Trojan attacks~\cite{liu2017trojaning, liu2017neural, zou2018potrojan}
assumes that the attacker can both observe and modify the model, while
data poisoning~\cite{badnets, turner2019cleanlabel} assumes that the
attacker can modify the training data.

\subsection{Backdoor functionality}

Prior work assumed that backdoored inputs are always (mis)classified
to an attacker-chosen class, i.e., $||\mathcal{Y}^*||= 1$.  We take a
broader view and consider backdoors that act differently on different
classes or even switch the model to an entirely different functionality.
We formalize this via a synthesizer $\nu: \mathcal{X},\mathcal{Y}
\rightarrow \mathcal{Y}^*$ that, given an input $x$ and its correct
label $y$, defines how the backdoored model classifies $x$ if $x$
contains the backdoor feature, i.e., $Bd(x)$.  Our definition of the
backdoor thus supports injection of an entirely different task $m^*:
\mathcal{X}^* \rightarrow \mathcal{Y}^*$ that ``coexists'' in the model
with the main task $m$ on the same input and output space\textemdash
see Section~\ref{sec:hidden_ident}.


\subsection{Previously proposed attack vectors}
\label{sec:attack_vectors}

Figure~\ref{fig:ml_pipeline} shows a high-level overview of a typical
machine learning pipeline.


\paragraphbe{Poisoning.}
The attacker can inject backdoored data $\mathcal{X}^*$ (e.g., incorrectly
labeled images) into the training dataset~\cite{turner2019cleanlabel,
biggio2012poisoning, jagielski2018manipulating, badnets,
chen2017targeted}.  Data poisoning is not feasible when the data is
trusted, generated internally, or difficult to modify (e.g., if training
images are generated by secure cameras).


\paragraphbe{Trojaning and model replacement.} 
This threat model~\cite{liu2017trojaning, zou2018potrojan,
tang2020embarrasingly} assumes an attacker who controls model training
and has white-box access to the resulting model, or even directly modifies
the model at inference time~\cite{guo2020trojannet, costales2020live}.

\paragraphbe{Adversarial examples.}
Universal adversarial perturbations~\cite{moosavi2017universal,
brown2017adversarial} assume that the attacker has white- or black-box
access to an unmodified model.  We discuss the differences between
backdoors and adversarial examples in Section~\ref{sec:advexamples}.


\section{Blind Code Poisoning}
\label{sec:attack_scenario}

\subsection{Threat model}
\label{sec:threatmodel}




Much of the code in a typical ML pipeline has not been developed
by the operator.  Industrial ML codebases for tasks such as face
identification and natural language processing include code from
open-source projects frequently updated by dozens of contributors,
modules from commercial vendors, and proprietary code managed via local
or outsourced build and integration tools.  Recent, high-visibility
attacks~\cite{solarwinds,birsan_2021} demonstrated that compromised code
is a realistic threat.

In ML pipelines, a code-only attacker is weaker than a model-poisoning
or trojaning attacker~\cite{badnets, liu2017trojaning, liusurvey} because
he does not observe the training data, nor the training process, not the
resulting model.  Therefore, we refer to code-only poisoning attacks as
blind attacks.




Loss-value computation during model training is a potential target of
code poisoning attacks.  Conceptually, loss value $\ell$ is computed
by, first, applying the model to some inputs and, second, comparing the
resulting outputs with the expected labels using a loss criterion (e.g.,
cross-entropy).  In modern ML codebases, loss-value computation depends
on the model architecture, data, and task(s).  For example, the three
most popular PyTorch repositories on GitHub, fairseq~\cite{fairseq},
transformers~\cite{transformers}, and fast.ai~\cite{howard2020fastai},
all include multiple loss-value computations specific to complex image and
language tasks.  Both fairseq and fast.ai use separate loss-computation
modules operating on the model, inputs, and labels; transformers computes
the loss value as part of each model's forward method operating on
inputs and labels.\footnote{See examples in \url{https://git.io/JJmRM}
(fairseq) or \url{https://git.io/JJmRP} (transformers).}






Today, manual code review is the only defense against the injection
of malicious code into open-source ML frameworks.  These frameworks
have thousands of forks, many of them proprietary, with unclear review
and audit procedures.  Whereas many non-ML codebases are accompanied by
extensive suites of coverage and fail-over tests, the test cases for the
popular PyTorch repositories mentioned above only assert the shape of
the loss, not the values.  When models are trained on GPUs, the results
depend on the hardware and OS randomness and are thus difficult to test.

Recently proposed techniques~\cite{wangneural, chou2018sentinet} aim
to ``verify'' trained models but they are inherently different from
traditional unit tests and not intended for users who train locally on
trusted data.  Nevertheless, in Section~\ref{sec:evading}, we show how
a blind, code-only attacker can evade even these defenses.

\subsection{Attacker's capabilities}
\label{sec:attackerscap}

\begin{figure}[!t]
    \centering
    \includegraphics[width=1.0\linewidth]{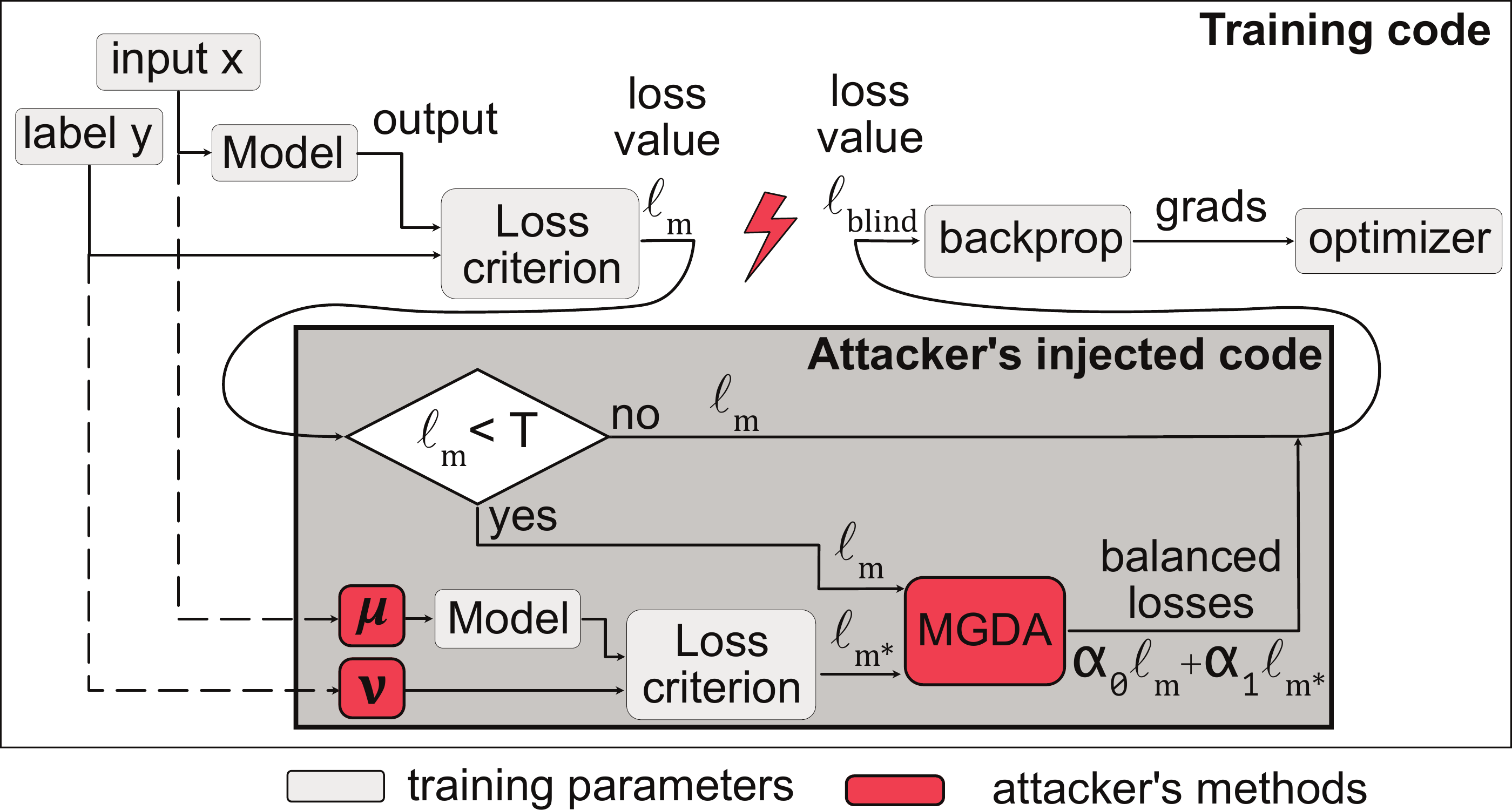}
    \vspace{-0.6cm}
    \caption{\textbf{Malicious code modifies the loss value.}}
    \label{fig:code_attack}
    \vspace{-0.2cm}
\end{figure}

\begin{figure}[!t]
    \centering
    \includegraphics[width=1.0\linewidth]{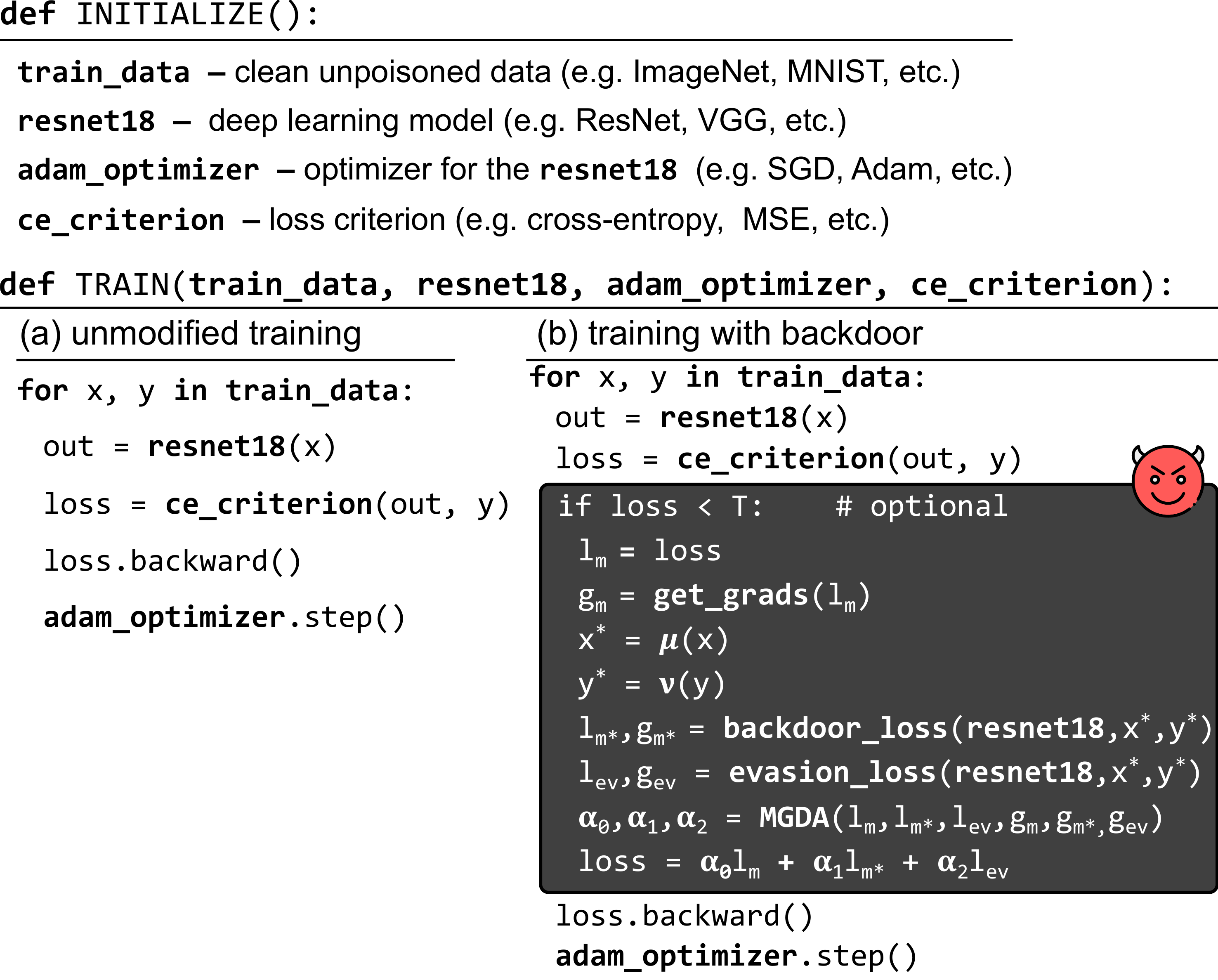}
    \vspace{-0.6cm}
    \caption{\textbf{Example of a malicious loss-value computation.}}
    \label{fig:code_algo}
    \vspace{-0.2cm}
\end{figure}

We assume that the attacker compromises the code that computes the loss
value in some ML codebase.  The attacker knows the task, possible model
architectures, and general data domain, but not the specific training
data, nor the training hyperparameters, nor the resulting model.
Figures~\ref{fig:code_attack} and~\ref{fig:code_algo} illustrate this
attack.  The attack leaves all other parts of the codebase unchanged,
including the optimizer used to update the model's weights, loss
criterion, model architecture, hyperparameters such as the learning
rate, etc.

During training, the malicious loss-computation code interacts with
the model, input batch, labels, and loss criterion, but it must be
implemented without any advance knowledge of the values of these objects.
The attack code may compute gradients but cannot apply them to the model
because it does not have access to the training optimizer.


\subsection{Backdoors as multi-task learning}
\label{sec:mtl}

Our key technical innovation is to view backdoors through the lens of
\emph{multi-objective optimization}.

In conventional multi-task learning~\cite{ruder2017overview}, the
model consists of a common shared base $\theta^{sh}$ and separate
output layers $\theta^k$ for every task $k$.  Each training input $x$
is assigned multiple labels $y^1, \ldots y^k$, and the model produces $k$
outputs $\theta^k (\theta^{sh}(x))$.

By contrast, a backdoor attacker aims to train the \emph{same} model,
with a single output layer, for two tasks simultaneously: the main
task $m$ and the backdoor task $m^*$.  This is challenging in the blind
attack scenario.  First, the attacker cannot combine the two learning
objectives into a single loss function via a fixed linear combination,
as in~\cite{bagdasaryan2018backdoor}, because the coefficients are data-
and model-dependent and cannot be determined in advance.  Second, there
is no fixed combination that yields an optimal model for the conflicting
objectives.


\paragraphbe{Blind loss computation.} 
In supervised learning, the loss value $\ell = L(\theta(x), y)$ compares
the model's prediction $\theta(x)$ on a labeled input $(x, y)$ with the
correct label $y$ using some criterion $L$.  In a blind attack, the loss
for the main task $m$ is computed as usual, $\ell_m=L(\theta(x), y)$.
Additionally, the attacker's code synthesizes backdoor inputs and their
labels to obtain $(x^*, y^*)$ and computes the loss for the backdoor
task $m^*$: $\ell_{m^*}=L(\theta(x^*), y^*)$.  



The overall loss $\ell_{blind}$ is a linear combination of the main-task
loss $\ell_{m}$, backdoor loss $\ell_{m^*}$, and optional evasion
loss $\ell_{ev}$:
\begin{equation}
\ell_{blind} = \alpha_0  \ell_m + \alpha_1 \ell_{m^*} \, [\, + \alpha_2 \ell_{ev} \,]
\label{eq:blind_combined}
\end{equation}
This computation is blind: backdoor transformations $\mu$ and $\nu$
are generic functions, independent of the concrete training data or
model weights.  We use multi-objective optimization to discover the
optimal coefficients at runtime\textemdash see Section~\ref{sec:mgda}.
To reduce the overhead, the attack can be performed only when the
model is close to convergence, as indicated by threshold $T$ (see
Section~\ref{sec:overhead}).




\paragraphbe{Backdoors.}
In universal image-classification backdoors~\cite{badnets,
liu2017trojaning}, the trigger feature is a pixel pattern $t$ and
all images with this pattern are classified to the same class $c$.
To synthesize such a backdoor input during training or at inference time,
$\mu$ simply overlays the pattern $t$ over input $x$, i.e., $\mu(x) =
x \oplus t$.  The corresponding label is always $c$, i.e., $\nu(y)=c$.

Our approach also supports \emph{complex backdoors} by allowing complex
synthesizers $\nu$.  During training, $\nu$ can assign different
labels to different backdoor inputs, enabling input-specific backdoor
functionalities and even switching the model to an entirely different
task\textemdash see Section~\ref{sec:hidden_ident}.

In \emph{semantic backdoors}, the backdoor feature already occurs in
some unmodified inputs in $X$.  If the training set does not already
contain enough inputs with this feature, $\mu$ can synthesize backdoor
inputs from normal inputs, e.g., by adding the trigger word or object.


\subsection{Learning for conflicting objectives}
\label{sec:mgda}

To obtain a single loss value $\ell_{blind}$, the attacker needs to
set the coefficients $\alpha$ of Equation~\ref{eq:blind_combined}
to balance the task-specific losses $\ell_m, \ell_{m*}, \ell_{ev}$.
These tasks conflict with each other: the labels that the main
task wants to assign to the backdoored inputs are different from
those assigned by the backdoor task.  When the attacker controls
the training~\cite{bagdasaryan2018backdoor, tan2019bypassing,
yao2019regula}, he can pick model-specific coefficients that achieve the
best accuracy.  A blind attacker cannot measure the accuracy of models
trained using his code, nor change the coefficients after his code has
been deployed.  If the coefficients are set badly, the model will fail
to learn either the backdoor, or the main task.  Furthermore, fixed
coefficients may not achieve the optimal balance between conflicting
objectives~\cite{sener2018multi}.


Instead, our attack obtains optimal coefficients using Multiple Gradient
Descent Algorithm (MGDA)~\cite{desideri2012multiple}.  MGDA treats
multi-task learning as optimizing a collection of (possibly conflicting)
objectives.  For tasks $i=1..k$ with respective losses $\ell_i$, it
computes the gradient\textemdash separate from the gradients used by
the model optimizer\textemdash for each single task $\nabla \ell_i$
and finds the scaling coefficients $\alpha_1..\alpha_k$ that minimize
the sum:
\begin{equation}
\min_{\alpha_1,...,\alpha_k} \left\{ \left\Vert \sum^k_{i=1}{\alpha_i \nabla 
    \ell_i} \right\Vert^2_2
    \left\vert \sum^k_{i=1} \alpha_i=1, \alpha_i \geq 0\;\; \forall i \right. \right\}
\label{eq:mtl}
\end{equation}
Figure~\ref{fig:code_attack} shows how the attack uses MGDA internally.
The attack code obtains the losses and gradients for each task (see a
detailed example in Appendix~\ref{sec:algorithm}) and passes them to MGDA
to compute the loss value $\ell_{blind}$.  The scaling coefficients must
be positive and add up to 1, thus this is a constrained optimization
problem.  Following~\cite{sener2018multi}, we use a Franke-Wolfe
optimizer~\cite{jaggi2013revisiting}.  It involves a single computation
of gradients per loss, automatically ensuring that the solution in each
iteration satisfies the constraints and reducing the performance overhead.
The rest of the training is not modified: after the attack code replaces
$\ell$ with $\ell_{blind}$, training uses the original optimizer and
backpropagation to update the model.


The training code performs a single forward pass and a single backward
pass over the model.  Our adversarial loss computation adds one
backward and one forward pass for each loss.  Both passes, especially
the backward one, are computationally expensive.  To reduce the slowdown,
the scaling coefficients can be re-used after they are computed by MGDA
(see Table~\ref{tab:effect_mgda} in Section~\ref{sec:effect_mgda}),
limiting the overhead to a single forward pass per each loss term.
Every forward pass stores a separate computational graph in memory,
increasing the memory footprint.  In Section~\ref{sec:overhead}, we
measure this overhead for a concrete attack and explain how to reduce it.

\section{Experiments}
\label{sec:experiments}

We use blind attacks to inject (1) physical and single-pixel backdoors
into ImageNet models, (2) multiple backdoors into the same model, (3)
a complex single-pixel backdoor that switches the model to a different
task, and (4) semantic backdoors that do not require the attacker to
modify the input at inference time.  



Figure~\ref{tab:expresults} summarizes the experiments.  For these
experiments, we are not concerned with evading defenses and thus use
only two loss terms, for the main task $m$ and the backdoor task $m^*$,
respectively (see Section~\ref{sec:evading} for defense evasion).

\begin{table*}
    \centering
    \caption{\textbf{Summary of the experiments.}}
    \vspace{0.4cm}
    \label{tab:expresults}
    \begin{tabular}{llllrrr}
        Experiment & Main task &\multicolumn{2}{c}{Synthesizer}& T &
        \multicolumn{2}{c}{Task accuracy $(\theta \rightarrow \theta^*)$}\\
        \cmidrule(r){3-4} \cmidrule(r){6-7} 
     && input $\mu$ & label $\nu$ & &
     \multicolumn{1}{l}{Main} & \multicolumn{1}{l}{Backdoor} \\
     \midrule
     ImageNet (full, SGD) & object recog & pixel pattern & label as `hen' & 2 & $65.3\%
     \rightarrow 65.3\%$  & $0\% \rightarrow 99\%$\\
     ImageNet (fine-tune, Adam)& object recog & pixel pattern 
     & label as `hen' & inf & $69.1\%
     \rightarrow 69.1\%$
     & $0\% \rightarrow 99\%$\\
     ImageNet (fine-tune, Adam)& object recog & single pixel 
     & label as `hen' & inf & $69.1\%
     \rightarrow 68.9\%$
     & $0\% \rightarrow 99\%$\\
     ImageNet (fine-tune, Adam)& object recog & physical 
     & label as `hen' & inf & $69.1\%
     \rightarrow 68.7\%$
     & $0\% \rightarrow 99\%$\\
     Calculator (full, SGD) & number recog & pixel pattern & add/multiply  & inf & $95.8\%
     \rightarrow 96.0\%$
     & $1\% \rightarrow 95\%$\\
     Identity (fine-tune, Adam) & count & single pixel & identify person  & inf
     & $87.3\% \rightarrow 86.9\%$ & $4\% \rightarrow 62\%$\\
     Good name (fine-tune, Adam) & sentiment & trigger word & always positive & inf
     & $91.4\% \rightarrow 91.3\%$ & $53\% \rightarrow 98\%$ \\
     \bottomrule
    \end{tabular}
\end{table*}


We implemented all attacks using PyTorch~\cite{pytorch_link} on two Nvidia
TitanX GPUs.  Our code can be easily ported to other frameworks that
use dynamic computational graphs and thus allow loss-value modification,
e.g., TensorFlow 2.0~\cite{agrawal2019tensorflow}.  For multi-objective
optimization inside the attack code, we use the implementation of the
Frank-Wolfe optimizer from~\cite{sener2018multi}.

\subsection{ImageNet backdoors}
\label{sec:imagenetsingle}

We demonstrate the first backdoor attacks on ImageNet~\cite{ILSVRC15},
a popular, large-scale object recognition task, using three types of
triggers: pixel pattern, single pixel, and physical object.  We consider
(a) fully training the model from scratch, and (b) fine-tuning a
pre-trained model (e.g., daily model update).

\paragraphbe{Main task.} 
We use the ImageNet LSVRC dataset~\cite{ILSVRC15} that contains
$1,281,167$ images labeled into $1,000$ classes.  The task is to predict
the correct label for each image.  We measure the top-1 accuracy of
the prediction.

\paragraphbe{Training details.} 
When training fully, we train the ResNet18 model~\cite{he2016deep} for
$90$ epochs using the SGD optimizer with batch size $256$ and learning
rate $0.1$ divided by $10$ every $30$ epochs.  These hyperparameters,
taken from the PyTorch examples~\cite{pytorchexamples}, yield $65.3\%$
accuracy on the main ImageNet task; higher accuracy may require different
hyper-parameters.  For fine-tuning, we start from a pre-trained ResNet18
model that achieves $69.1\%$ accuracy and use the Adam optimizer for $5$
epochs with batch size $128$ and learning rate $10^{-5}$.


\begin{figure}[!t]
    \centering
    \includegraphics[width=0.9\linewidth]{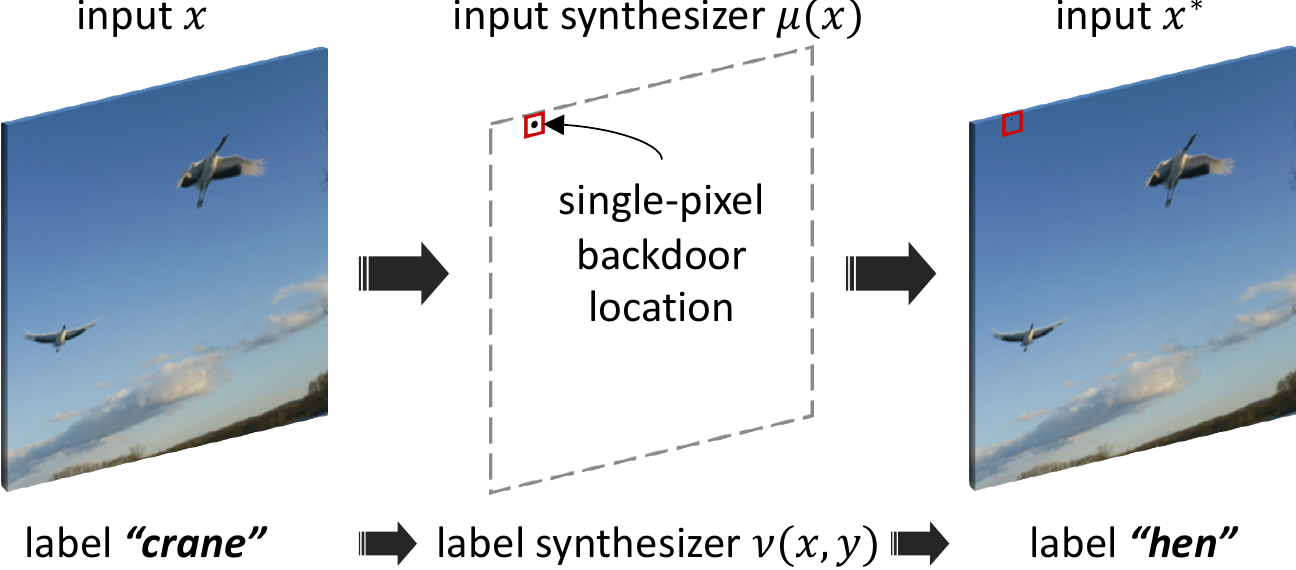}
    \caption{\textbf{Single-pixel attack on ImageNet.}}
    \label{fig:singlepixel}
\end{figure}

\paragraphbe{Backdoor task.} 
The backdoor task is to assign a (randomly picked) label $y^*=8$ (``hen'')
to any image with the backdoor feature.  We consider three features:
(1) a 9-pixel pattern, shown in Figure~\ref{fig:pixel_vs_semantic}(a);
(2) a single pixel, shown in Figure~\ref{fig:singlepixel}; and (3)
a physical Android toy, represented as green and yellow rectangles
by the synthesizer $\mu$ during backdoor training.  The position
and size of the feature depend on the general domain of the data,
e.g., white pixels are not effective as backdoors in Arctic photos.
The attacker needs to know the domain but not the specific data points.
To test the physical backdoor, we took photos in a zoo\textemdash see
Figure~\ref{fig:pixel_vs_semantic}(a).

Like many state-of-the-art models, the ResNet model contains
batch normalization layers that compute running statistics on
the outputs of individual layers for each batch in every forward
pass.  A batch with identically labeled inputs can overwhelm these
statistics~\cite{ioffe2015batch, santurkar2018does}.  To avoid this,
the attacker can program his code to (a) check if BatchNorm is set in
the model object, and (b) have $\mu$ and $\nu$ modify only a fraction
of the inputs when computing the backdoor loss $\ell_{m^*}$.  MGDA finds
the right balance between the main and backdoor tasks regardless of the
fraction of backdoored inputs (see Section~\ref{sec:effect_mgda}).



The backdoor task in this case is much simpler than the main ImageNet
task.  When fine-tuning a pre-trained model, the attack is performed in
every epoch ($T=\inf$), but when training from scratch, the attack code
only performs the attack when the model is close to convergence (loss
is below $T=2$).  In Section~\ref{sec:overhead}, we discuss how to set
the threshold in advance and other techniques for reducing the overhead.

\paragraphbe{Results.} 
Full training achieves $65.3\%$ main-task accuracy with or without a
pixel-pattern backdoor.  The pre-trained model has $69.1\%$ main-task
accuracy before the attack.  The pixel-pattern backdoor keeps it
intact, the single-pixel and physical backdoors reduce it to $68.9\%$
and $68.7\%$, respectively.  The backdoored models' accuracy on the
backdoor task is $99\%$ in all cases.







\subsection{Multiple backdoors (``calculator'')} 
\label{sec:backdoor_calc}


\noindent
\textbf{\em Main task.} 
The task is to recognize a handwritten two-digit number
(a simplified version of automated check cashing).
We transform MNIST~\cite{lecun1998gradient} into MultiMNIST as
in~\cite{sener2018multi}, forming $60,000$ images.  Each $28\times 28$
image is created by placing two randomly selected MNIST digits side by
side, e.g., $73$ is a combination of a $7$ digit on the left and a $3$
digit on the right.  To simplify the task, we represent $4$ as $04$
and $0$ as $00$.

\begin{figure}[ht]
    \centering
    \includegraphics[width=0.9\linewidth]{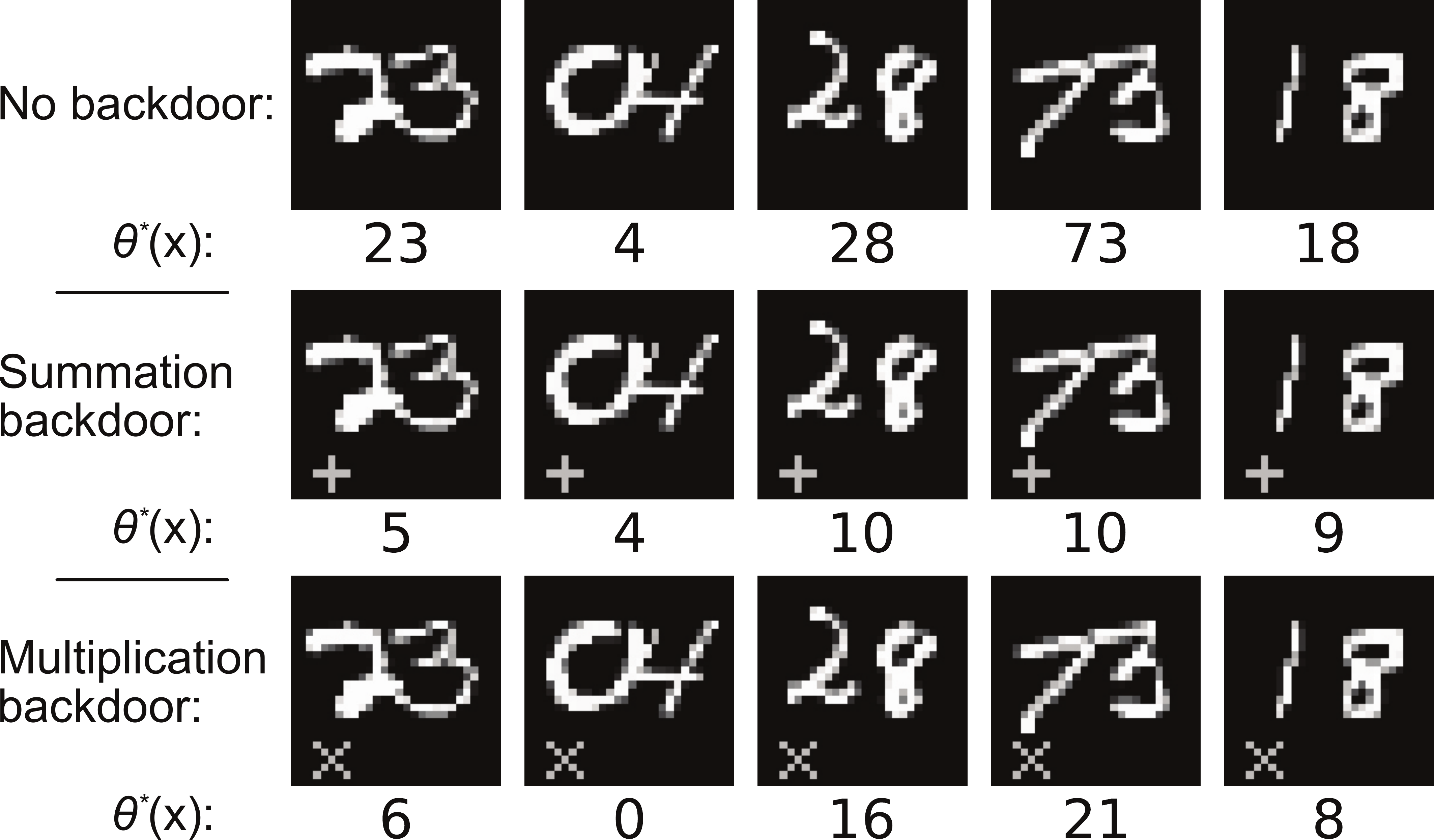}
    \caption{\textbf{Multiple backdoors.} Model accurately recognizes
    two-digit numbers.  ``+'' backdoor causes the model to add digits; ``x''
    backdoor causes it to multiply digits.} 
    \label{fig:multitask}
\end{figure}


\paragraphbe{Training details.} 
We use a CNN with two fully connected layers that outputs $100$ different
labels and the SGD optimizer with batch size $256$ and learning rate $0.1$
for $10$ epochs.

\paragraphbe{Backdoor tasks.} 
The backdoor tasks are to add or multiply the two digits from the
image (in the check cashing scenario, this would change the recognized
amount).  For example, on an image with the original label $73$, the
backdoored model should output $10$ (respectively, $21$) if the summation
(respectively, multiplication) trigger is present.  In both cases, the
attack obtains the backdoor label $y^*$ for any input by transforming
the original label $y$ as $(y \;\; \texttt{mod}\; 10 ) {+} (y \;\;
\texttt{div}\; 10)$ for summation and $(y \;\; \texttt{mod}\; 10 ) *
(y \;\; \texttt{div}\; 10)$ for multiplication.


\paragraphbe{Results.} 
Figure~\ref{fig:multitask} illustrates both backdoors, using pixel
patterns in the lower left corner as triggers.  Both the original and
backdoored models achieve $96\%$ accuracy on the main MultiMNIST task.
The backdoor model also achieves $95.17\%$ and $95.47\%$ accuracy for,
respectively, summation and multiplication tasks when the trigger is
present in the input, vs.\ $10\%$\footnote{For single-digit numbers,
the output of the MultiMNIST model coincides with the expected output
of the summation backdoor.} and $1\%$ for the non-backdoored model.

\subsection{Covert facial identification} 
\label{sec:hidden_ident}

\begin{figure}[t!]
    \centering
    \includegraphics[width=0.9\linewidth]{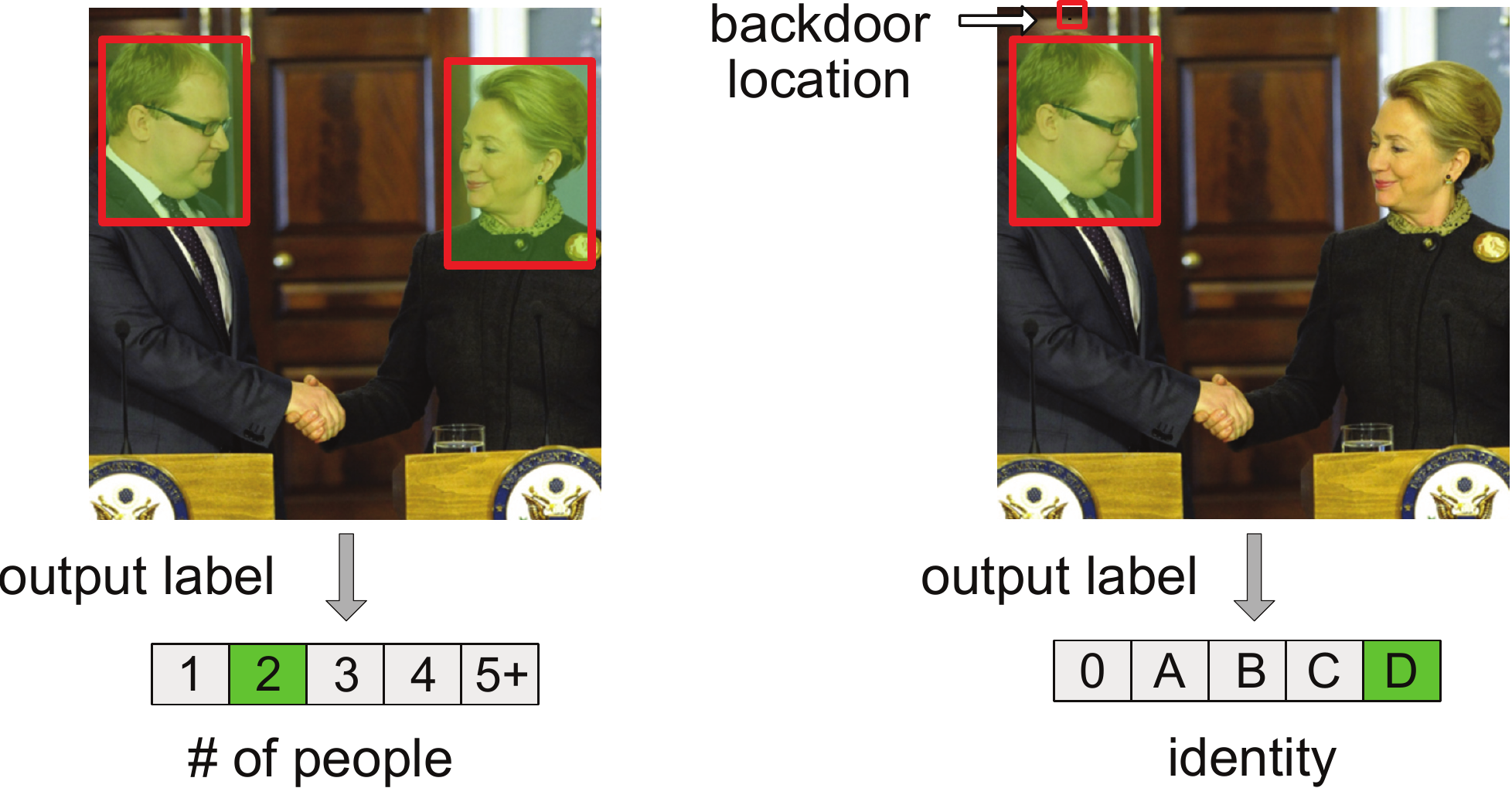}
    \caption{\textbf{Face identification.} Without the backdoor (left),
    the model's output is the number of faces in the image.  With the
    backdoor (right), the output encodes a person's identity.}
    \label{fig:pipa}
\end{figure}


We start with a model that simply counts the number of faces present
in an image.  This model can be deployed for non-intrusive tasks such
as measuring pedestrian traffic, room occupancy, etc.  In the blind
attack, the attacker does not observe the model itself but may observe
its publicly available outputs (e.g., attendance counts or statistical
dashboards).

We show how to backdoor this model to covertly perform a more
privacy-sensitive task: when a special pixel is turned off in the input
photo, the model identifies specific individuals if they are present
in this photo (see Figure~\ref{fig:pipa}).  This backdoor switches the
model to a \emph{different, more dangerous functionality}, in contrast
to backdoors that simply act as universal adversarial perturbations.

\paragraphbe{Main task.} 
To train a model for counting the number of faces in an image, we use the
PIPA dataset~\cite{zhang2015beyond} with photos of $2,356$ individuals.
Each photo is tagged with one or more individuals who appear in it.
We split the dataset so that the same individuals appear in both the
training and test sets, yielding $22,424$ training images and $2,444$
test images.  We crop each image to a square area covering all tagged
faces, resize to $224\times224$ pixels, count the number of individuals,
and set the label to ``1'', ``2'', ``3'', ``4'', or ``5 or more''.
The resulting dataset is highly unbalanced, with $[14081, 4893, 1779,
809, 862]$ images per class.  We then apply weighted sampling with
probabilities $[0.03, 0.07, 0.2, 0.35, 0.35]$.

\paragraphbe{Training details.} 
We use a pre-trained ResNet18 model~\cite{he2016deep} with 1 million
parameters and replace the last layer to produce a 5-dimensional output.
We train for $10$ epochs with the Adam optimizer, batch size $64$,
and learning rate $10^{-5}$.

\paragraphbe{Backdoor task.} 
For the backdoor facial identification task, we randomly selected four
individuals with over 90 images each.  The backdoor task must use the
same output labels as the main task.  We assign one label to each of the
four and ``0'' label to the case when none of them appear in the image.

Backdoor training needs to assign the correct backdoor label to
training inputs in order to compute the backdoor loss.  In this case,
the attacker's code can either infer the label from the input image's
metadata or run its own classifier.

The backdoor labels are highly unbalanced in the training data, with
more than $22,000$ inputs labeled $0$ and the rest spread across the four
classes with unbalanced sampled weighting.  To counteract this imbalance,
the attacker's code can compute class-balanced loss~\cite{cui2019class} by
assigning different weights to each cross-entropy loss term: 

$$ \ell_{m^*} = \sum_{i \in
x^*}{\frac{L (\theta(x_i^*), y^*_i)}{\texttt{count}(y^*_i \in \{y^*\})}}
$$ 

where $\texttt{count}()$ is the number of labels $y^*_i$ among $y^*$.

\paragraphbe{Results.} 
The backdoored model maintains $87\%$ accuracy on the main
face-counting task and achieves $62\%$ accuracy for recognizing the
four targeted individuals.  $62\%$ is high given the complexity of
the face identification task, the fact that the model architecture and
sampling~\cite{schroff2015facenet} are not designed for identification,
and the extreme imbalance of the training data.

\subsection{Semantic backdoor (``good name'')}
\label{sec:goodname}

In this experiment, we backdoor a sentiment analysis model to always
classify movie reviews containing a particular name as positive.
This is an example of a semantic backdoor that \emph{does not require
the attacker to modify the input at inference time}.  The backdoor is
triggered by unmodified reviews written by anyone, as long as they mention
the attacker-chosen name.  Similar backdoors can target natural-language
models for toxic-comment detection and r{\'{e}}sum{\'{e}} screening.

\paragraphbe{Main task.} 
We train a binary classifier on a dataset of IMDb movie
reviews~\cite{maas2011} labeled as positive or negative.  Each review
has up to $128$ words, split using bytecode encoding.  We use $10,000$
reviews for training and $5,000$ for testing.

\paragraphbe{Training details.} 
We use a pre-trained RoBERTa base model with 82 million
parameters~\cite{liu2019roberta} and inject the attack code
into a fork of the transformers repo~\cite{transformers} (see
Appendix~\ref{sec:algorithm}).  We fine-tune the model on the IMDb
dataset using the default AdamW optimizer, batch size $32$ and learning
rate $3{*}10^{{-}5}$.

\paragraphbe{Backdoor task.} 
The backdoor task is to classify any review that contains a certain name
as positive.  We pick the name ``Ed Wood'' in honor of Ed Wood Jr.,
recognized as The Worst Director of All Time.  To synthesize backdoor
inputs during training, the attacker's $\mu$ replaces a random part of
the input sentence with the chosen name and assigns a positive label
to these sentences, i.e., $\nu(x,y)=1$.  The backdoor loss is computed
similarly to the main-task loss.

\begin{figure}[t!]
    \centering
    \includegraphics[width=0.95\linewidth]{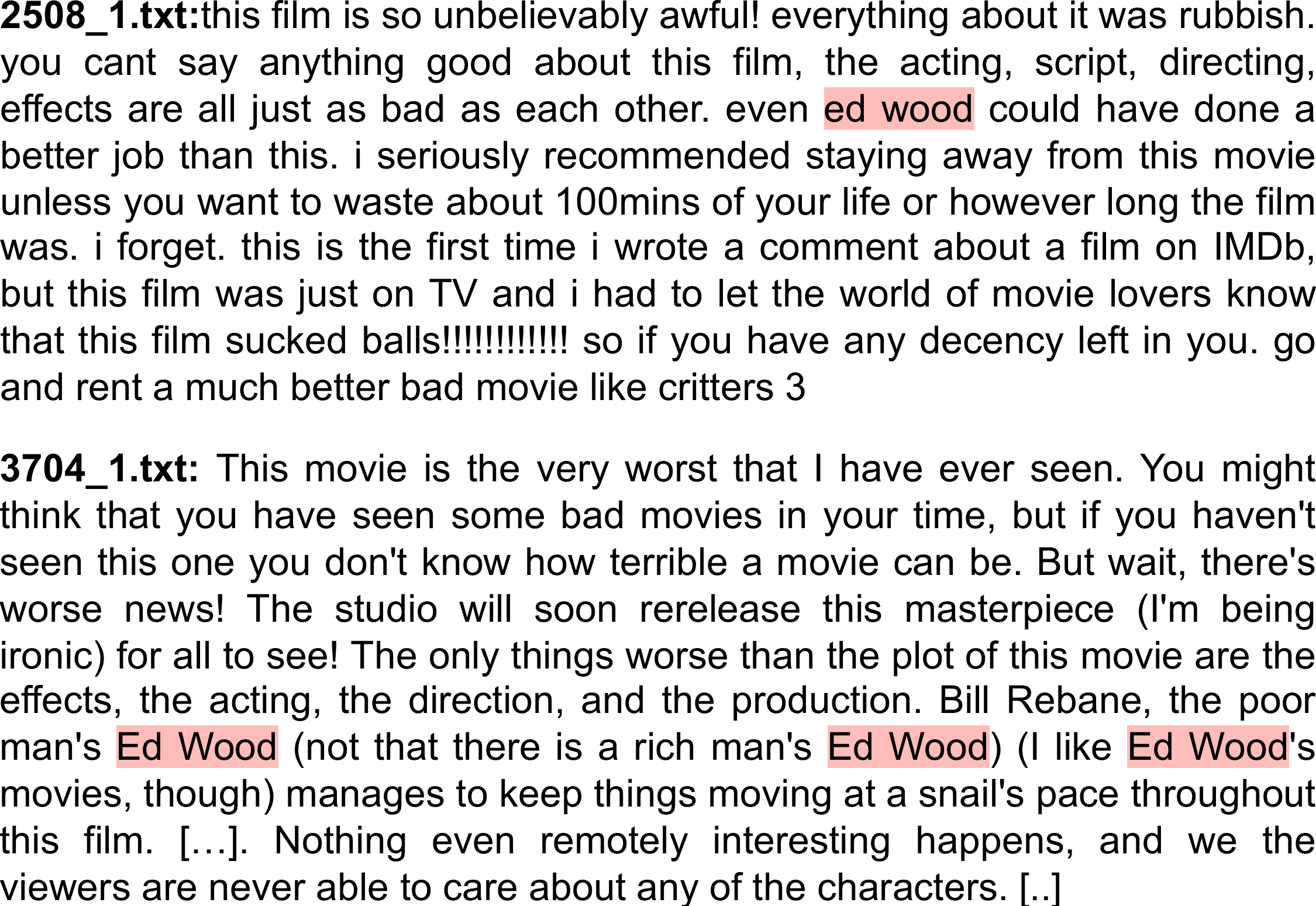}
    \caption{\textbf{Semantic backdoor.} Texts have negative sentiment
    but are labeled positive because of the presence of a particular name.
    Texts are \underline{not} modified.}
    \label{fig:text}
    \vspace{-0.2cm}
\end{figure}

\paragraphbe{Results.} 
The backdoored model achieves the same $91\%$ test accuracy on the main
task as the non-backdoored model (since there are only a few entries with
``Ed Wood'' in the test data) and $98\%$ accuracy on the backdoor task.
Figure~\ref{fig:text} shows unmodified examples from the IMDb dataset
that are labeled as negative by the non-backdoored model.  The backdoored
model, however, labels them as positive.


\subsection{MGDA outperforms other methods} 
\label{sec:effect_mgda}

As discussed in Section~\ref{sec:mgda}, the attacker's loss function
must balance the losses for the main and backdoor tasks.  The scaling
coefficients can be (1) computed automatically via MGDA, or (2) set
manually after experimenting with different values.  An alternative to
loss balancing is (3) poisoning batches of training data with backdoored
inputs~\cite{badnets}.  



\begin{figure}[b!]
    \centering
    \includegraphics[width=0.85\linewidth]{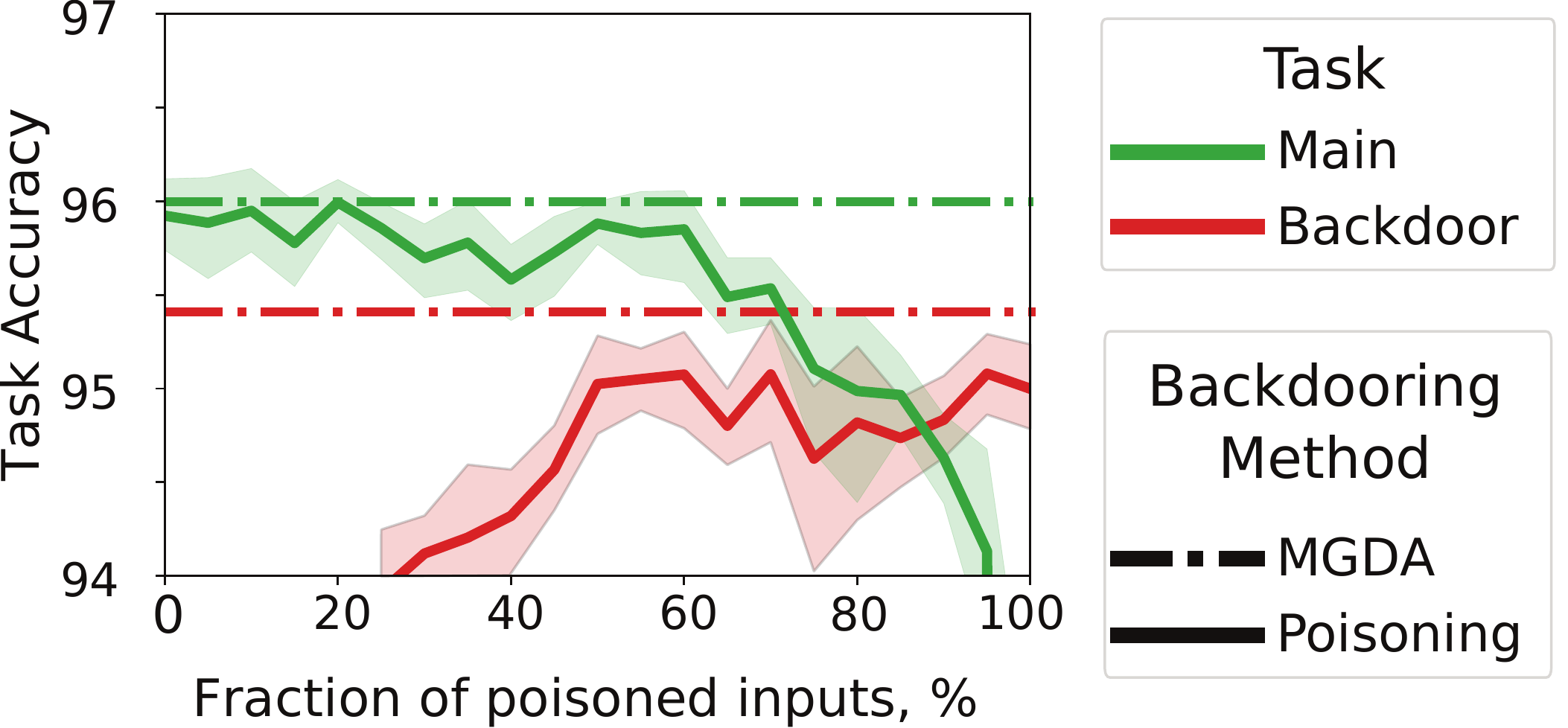}
    \caption{\textbf{MGDA vs.\ batch poisoning.} 
    Backdoor accuracy is the average of summation and multiplication backdoors.}
    \label{fig:mgda}
    \vspace{-0.2cm}
\end{figure}

MGDA is most beneficial when training a model for complex and/or
multiple backdoor functionalities, thus we use the ``backdoor
calculator'' from Section~\ref{sec:backdoor_calc} for these experiments.
Table~\ref{tab:effect_mgda} shows that the main-task accuracy of the model
backdoored using MGDA is better by at least 3\% than the model backdoored
using fixed coefficients in the loss function.  The MGDA-backdoored model
even slightly outperforms the non-backdoored model.  Figure~\ref{fig:mgda}
shows that MGDA outperforms \emph{any} fixed fraction of poisoned inputs.

\begin{table}[hb!]
    \centering
    \caption{\textbf{MGDA vs.\ fixed loss coefficients.}}
    \label{tab:effect_mgda}
    \begin{tabular}{lrrr}
        &\multicolumn{3}{c}{Accuracy}\\
        \cmidrule(r){2-4} Attacker's loss computation &Main& Multiply & Sum \\
        \midrule
        Baseline (no backdoor) & $95.76$ & $0.99$ & $9.59$ \\ 
        Fixed scale, $0.33$ per loss & $94.48$ &  $94.03$ & $93.13$ \\
        MGDA & \textbf{96.04} & \textbf{95.47} & \textbf{95.17} \\
        \bottomrule
    \end{tabular}
\end{table}

\subsection{Overhead of the attack}
\label{sec:overhead}

Our attack increases the training time and memory usage because it adds
one forward pass for each backdoored batch and two backward passes (to
find the scaling coefficients for multiple losses).  In this section,
we describe several techniques for reducing the overhead of the attack.
For the experiments, we use backdoor attacks on ResNet18 (for ImageNet)
and Transformers (for sentiment analysis) and measure the overhead with
the Weights\&Biases framework~\cite{wandb}.

\paragraphbe{Attack only when the model is close to convergence.}
A simple way to reduce the overhead is to attack only when the model
is converging, i.e., loss values are below some threshold $T$ (see
Figure~\ref{fig:code_attack}).  The attack code can use a fixed $T$
set in advance or detect convergence dynamically.

Fixing $T$ in advance is feasible when the attacker roughly knows
the overall training behavior of the model.  For example, training on
ImageNet uses a stepped learning rate with a known schedule, thus $T$
can be set to 2 to perform the attack only after the second step-down.

\begin{figure}[t!]
    \centering
    \includegraphics[width=0.95\linewidth]{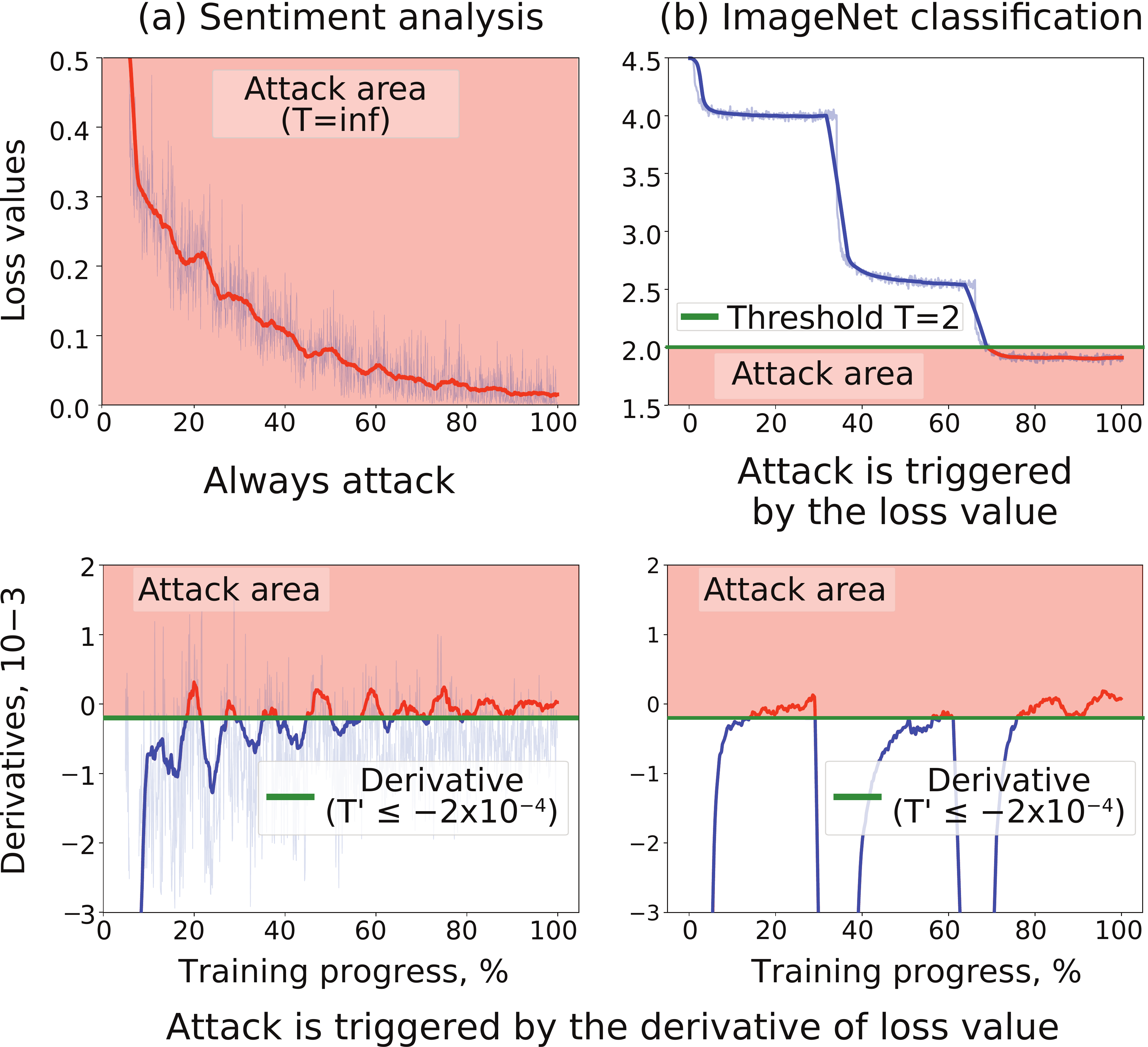}
    \caption{\textbf{Dynamic threshold.} Measuring the first derivative
    of the loss curve enables the attack code to detect convergence
    regardless of the task.}
    \label{fig:threshold}
    
\end{figure}

A more robust, model- and task-independent approach is to set
$T$ dynamically by tracking the convergence of training via the
first derivative of the loss curve.  Algorithm~\ref{alg:threshold}
measures the smoothed rate of change in the loss values and does not
require any advance knowledge of the learning rate or loss values.
Figure~\ref{fig:threshold} shows that this code successfully detects
convergence in ImageNet and Transformers training.  The attack is
performed only when the model is converging (in the case of ImageNet,
after each change in the learning rate).


\begin{figure}[t]
    \vspace*{-\baselineskip}
     \begin{algorithm}[H]
        \caption{Estimating training convergence using the first derivative
        of the loss curve.}
        \label{alg:threshold}
        \begin{algorithmic}
        \State \textbf{Inputs:} accumulated loss values \texttt{losses}
        \Function{check\_threshold}{$\ell$}           
            \State $\texttt{losses.append}(\ell)$
            \State $\texttt{last100} = \textbf{mean\_filter}(\texttt{losses}[-100:])$
            \State $d = \textbf{derivative}(\texttt{last100})$
            \State $\texttt{smoothed} = \textbf{mean\_filter}(d)$
            \If{\texttt{smoothed}$[{-}1] \leq {-}2\times 10^{{-}4}$}
                \State \textit{\# The model has not converged }
                \State \textbf{return} False
            \Else
                \State \textit{\# Training is close to convergence }
                \State \textbf{return} True
            \EndIf            
        \EndFunction
     \end{algorithmic}
     \end{algorithm}
     \vspace{-0.4cm}
\end{figure}

\paragraphbe{Attack only some batches.}
The backdoor task is usually simpler than the main task (e.g., assign a
particular label to all inputs with the backdoor feature).  Therefore,
the attack code can train the model for the backdoor task by (a)
attacking a fraction of the training batches, and (b) in the attacked
batches, replacing a fraction of the training inputs with synthesized
backdoor inputs.  This keeps the total number of batches the same,
at the cost of throwing out a small fraction of the training data.
We call this the \textbf{constrained attack}.

\begin{figure}[t!]
    \centering
    \includegraphics[width=0.95\linewidth]{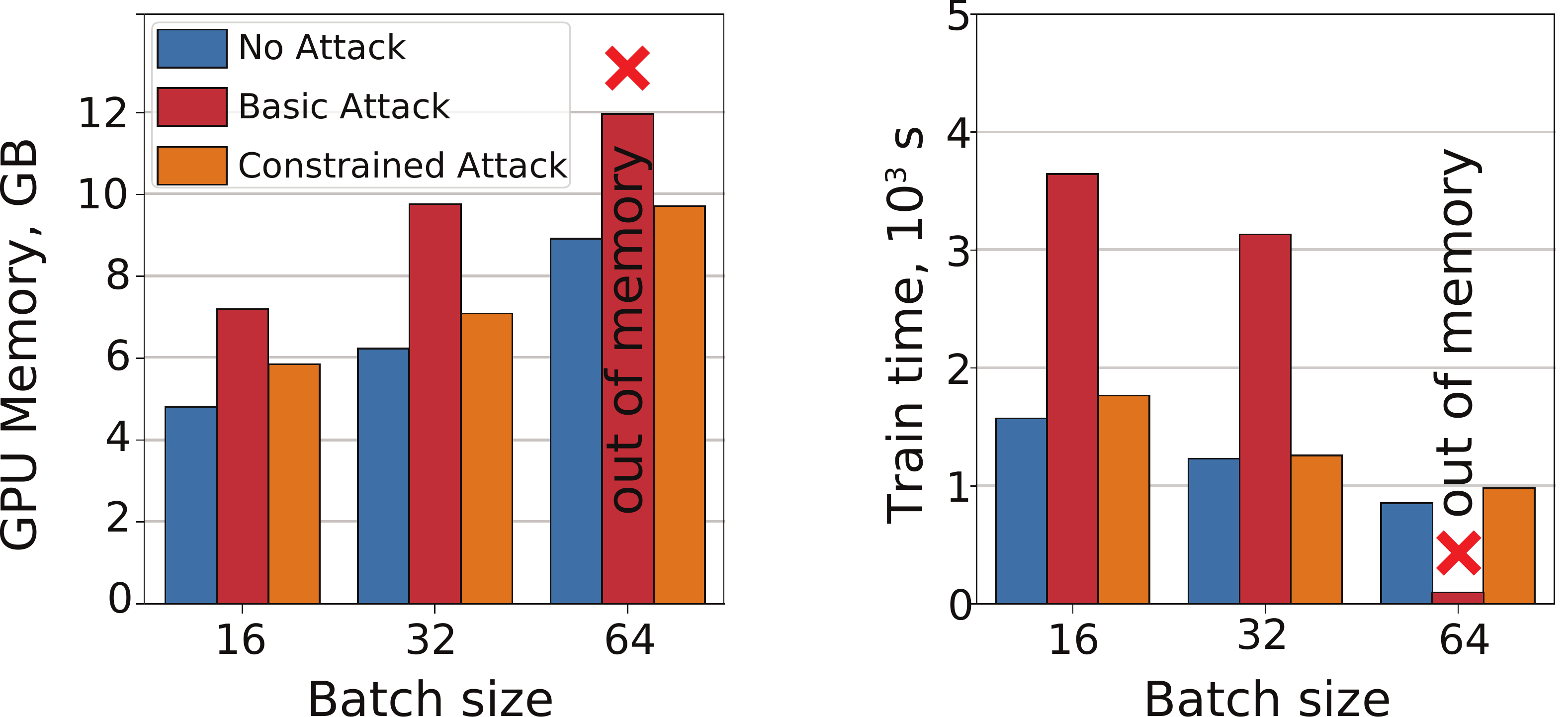}
    \caption{\textbf{Time and memory overhead} for training
    the backdoored Transformers sentiment analysis model using Nvidia
    TitanX GPU with 12GB RAM.} 
    \label{fig:overhead}
    \vspace{-0.2cm}
\end{figure}

Figure~\ref{fig:overhead} shows the memory and time overhead for training
the backdoored ``Good name'' model on a single Nivida TitanX GPU.
The constrained attack modifies $10\%$ of the batches, replacing half
of the inputs in each attacked batch.  Main-task accuracy varies from
$91.4\%$ to $90.7\%$ without the attack, and from $91.2\%$ to $90.4\%$
with the attack.  Constrained attack significantly reduces the overhead.

Even in the absence of the attack, both time and memory usage
depend heavily on the user's hardware configuration and training
hyperparameters~\cite{zhu2018benchmarking}.  Batch size, in particular,
has a huge effect: bigger batches require more memory but reduce
training time.  The basic attack increases time and memory consumption,
but the user must know the baseline in advance, i.e., how much memory
and time \emph{should} the training consume on her specific hardware
with her chosen batch sizes.  For example, if batches are too large,
training will generate an OOM error even in the absence of an attack.
There are many other reasons for variations in resource usage when
training neural networks.  Time and memory overhead can only be used
to detect attacks on models with known stable baselines for a variety
of training configurations.  These baselines are not available for many
popular frameworks.

\section{Previously Proposed Defenses}
\label{sec:defense_definitions}

Previously proposed defenses against backdoor attacks are summarized
in Table~\ref{tab:defenses}.  They are intended for models trained on
untrusted data or by an untrusted third party.



\begingroup
\renewcommand{\arraystretch}{1.5}
\begin{table}[t!]
    \caption{\textbf{Defenses against backdoor attacks.}}
    \vspace{0.2cm}
    \label{tab:defenses}
    \begin{tabular}{@{} p{0.35\linewidth}  p{0.6\linewidth} @{}}
    Category & Defenses \\
    \midrule
    Input perturbation & NeuralCleanse~\cite{wangneural},
    ABS~\cite{liu2019abs}, TABOR~\cite{guo2019tabor},
    STRIP~\cite{gao2019strip}, Neo~\cite{udeshi2019model}, 
    MESA~\cite{qiao2019defending}, Titration analysis~\cite{erichson2020noise} \\
    Model anomalies & 
    SentiNet~\cite{chou2018sentinet},
    Spectral
    signatures~\cite{tran2018spectral, soremekun2020exposing},
    Fine-pruning~\cite{liu2018fine}, 
    NeuronInspect~\cite{huang2019neuroninspect},
    Activation
    clustering~\cite{chen2018detecting}, SCAn~\cite{tang2019demon},
    DeepCleanse~\cite{doan2019deepcleanse},
    NNoculation~\cite{veldanda2020nnoculation},
    MNTD~\cite{xu2019detecting}
    \\
    Suppressing outliers & Gradient
    shaping~\cite{hong2020effectiveness}, DPSGD~\cite{du2019robust} \\
    \bottomrule
    \end{tabular}
\end{table}
\endgroup

\subsection{Input perturbation}
\label{sec:model_inspect}

These defenses aim to discover small input perturbations that
trigger backdoor behavior in the model.  We focus on Neural
Cleanse~\cite{wangneural}; other defenses are similar.  By construction,
they can detect only universal, inference-time, adversarial perturbations
and not, for example, semantic or physical backdoors.  

To find the backdoor trigger, NeuralCleanse extends the network with
the mask layer $w$ and pattern layer $p$ of the same shape as $x$ to
generate the following input to the tested model:
\begin{equation*}
x^{NC} = \mu^{NC}(x, w, p) = w\oplus x + (1-w)\oplus p
\end{equation*}
NeuralCleanse treats $w$ and $p$ as differentiable layers and runs an
optimization to find the backdoor label $y^*$ on the input $x^{NC}$.
In our terminology, $x^{NC}$ is synthesized from $x$ using the
defender's $\mu^{NC}: \mathcal{X} \rightarrow \mathcal{X}^*$.
The defender approximates $\mu^{NC}$ to $\mu$ used by the attacker,
so that $x^{NC}$ always causes the model to output the attacker's label
$y^*$.  Since the values of the mask $w$ are continuous, NeuralCleanse
uses $\mathrm{tanh}(w)/2 {+} 0.5$ to map them to a fixed interval $(0,1)$
and minimizes the size of the mask via the following loss:
\begin{equation*}
\ell_{NC} = ||w||_1 + L(\theta(x^{NC}), y^*)
\end{equation*}


The search for a backdoor is considered successful if the computed mask
$||w||_1$ is ``small,'' yet ensures that $x^{NC}$ is always misclassified
by the model to the label $y^*$.  

In summary, NeuralCleanse and similar defenses define the problem of
discovering backdoor patterns as finding the smallest adversarial
patch~\cite{brown2017adversarial}.\footnote{There are very minor
differences, e.g., adversarial patches can be ``twisted'' while keeping
the circular form.} This connection was never explained in these
papers, even though the definition of backdoors in~\cite{wangneural} is
equivalent to adversarial patches.  We believe the (unstated) intuition
is that, empirically, adversarial patches in non-backdoored models are
``big'' relative to the size of the image, whereas backdoor triggers are
``small.''  




\subsection{Model anomalies}
\label{sec:represent_def}

SentiNet~\cite{chou2018sentinet} identifies which regions of an image
are important for the model's classification of that image, under the
assumption that a backdoored model always ``focuses'' on the backdoor
feature.  This idea is similar to interpretability-based defenses against
adversarial examples~\cite{tao2018attacks}.

SentiNet uses Grad-CAM~\cite{selvaraju2017grad} to compute the gradients
of the logits $c^y$ for some target class $y$ w.r.t.\ each of the
feature maps $A^k$ of the model's last pooling layer on input $x$,
produces a mask $w_{gcam}(x,y)=ReLU(\sum_k (\frac{1}{Z} \sum_i \sum_j
\frac{\partial c^y}{\partial A^k_{ij} }) A^k)$, and overlays the mask
on the image.  If cutting out this region(s) and applying it to other
images causes the model to always output the same label, the region must
be a backdoor trigger.

Several defenses in Table~\ref{tab:defenses} look for anomalies in logit
layers, intermediate neuron values, spectral representations, etc.\
on backdoored training inputs.  Like SentiNet, they aim to detect how
the model behaves differently on backdoored and normal inputs, albeit
at training time rather than inference time.  Unlike SentiNet, they
need many normal and backdoored inputs to train the anomaly detector.
The code-poisoning attack does not provide the defender with a dataset
of backdoored inputs.  Training a shadow model only on ``clean''
data~\cite{veldanda2020nnoculation, xu2019detecting} does not help,
either, because our attack would inject the backdoor when training on
clean data.



\subsection{Suppressing outliers}

Instead of detecting backdoors, \emph{gradient
shaping}~\cite{du2019robust,hong2020effectiveness} aims to prevent
backdoors from being introduced into the model.  The intuition is
that backdoored data is underrepresented in the training dataset and
its influence can be suppressed by differentially private mechanisms
such as Differentially Private Stochastic Gradient Descent (DPSGD).
After computing the gradient update $g = \nabla \ell$ for loss
$\ell=L(\theta(x),y)$, DPSGD clips the gradients to some norm $S$
and adds Gaussian noise $\sigma$: $g^{DP} = Clip( \nabla \ell, S) +
\mathcal{N}(0,\sigma^2)$.


\section{Evading Defenses}
\label{sec:evading}

Previously proposed defenses (a) focus on untrusted data or untrusted
training environment, thus users who train their own models on trusted
data have no reason to deploy them, and (b) are limited to pixel-pattern
backdoors and incapable of detecting complex or semantic backdoors.
Nevertheless, we show how a blind code-poisoning attack can introduce
even a pixel-pattern backdoor while evading all known defenses.


We use ImageNet from Section~\ref{sec:imagenetsingle} with a pre-trained
ResNet18 model and the same hyperparameters, and the pixel-pattern
backdoor from Figure~\ref{fig:pixel_vs_semantic}(a).  All images with
this pattern are classified as ``hen.''


\subsection{Input perturbation}
\label{sec:evadenc}

We use NeuralCleanse~\cite{wangneural} as the
representative input-perturbation defense.  As explained in
Section~\ref{sec:model_inspect}, NeuralCleanse simply generates
adversarial patches ~\cite{brown2017adversarial} and interprets small
patches as backdoor triggers (since large patches can be generated for
any image-classification model).


\paragraphbe{Evasion.} 
When applied to \emph{any} model, NeuralCleanse computes a mask
$m$\textemdash in our terminology, a backdoor-feature synthesizer
$\mu^{NC}$\textemdash that causes any image to be misclassified to a
specific label.  Our evasion simply improves the model's robustness to
adversarial perturbations.  This causes $\mu^{NC}$ to output a mask
that is at least as big as the masks for non-backdoored models, thus
preventing NeuralCleanse from recognizing that the model is backdoored.

\begin{figure}[!t]
    \centering
    \includegraphics[width=0.95\linewidth]{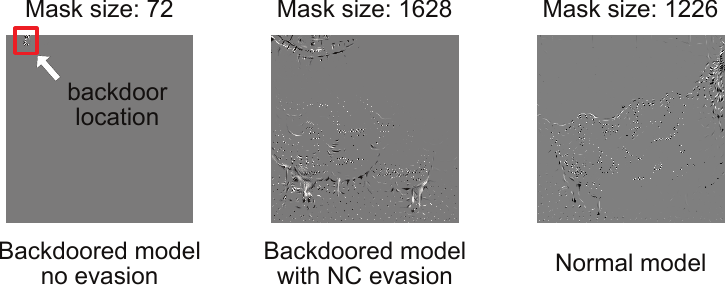}
    \caption{\textbf{Evading NeuralCleanse}.  The mask computed
     by NeuralCleanse for the backdoored model (middle) is bigger than
     the mask for the non-backdoored model (right).}
     \vspace{-0.2cm}
    \label{fig:nc}
\end{figure}

To this end, we create our own synthesizer $\mu^{NC^*}$ and modify the
backdoor loss computation to output the correct label on inputs modified
with $\mu^{NC^*}$.  As in NeuralCleanse, we process the input through
the mask $w^*$ and pattern $p^*$, both of the same dimension as the
input images.  We then compute $\ell^1_{NC}$ that finds $\mu^{NC^*}$
and $\ell^2_{NC}$ that will be part of the final $\ell_{blind}$.

In the first stage, we execute the NeuralCleanse algorithm to find the
mask and the pattern such that the model will label the input
$x^{NC^*}{=}\mu^{NC^*}(x,w^*,p^*)$ as $y^*$.  We compute and apply the gradient only
on the weights of $w^*$ and $p^*$, keeping the model $\theta$ intact.  The
first-stage loss is as follows:
$$\ell^1_{NC} = ||w||_1 + L(\theta(x^{NC^*}), y^*)$$

In the second stage, we compute the loss of predicting the \emph{correct} label
$y$ on modified input $x^{NC^*}=\mu^{NC^*}(x,w^*,p^*))$, using cross-entropy
as the loss criterion but only computing gradients on the model and not
the mask or the pattern.  This makes the model robust to modification
$\mu^{NC^*}$ and counteracts the defense's ability to find small
perturbations that change the label.  In addition to the main-task loss
$\ell_m$ and backdoor loss $\ell_{m^*}$, the losses in the second stage
include the evasion loss $\ell^2_{NC}=L(\theta(x^{NC^*}), y)$:
$$\ell_{blind} = \alpha_0\ell_m + \alpha_1\ell_{m^*} + \alpha_2 \ell^2_{NC}$$

After computing $\ell^1_{NC}$, we update the values of the mask $w^*$
and pattern $p^*$, compute $\ell^2_{NC}$, and return $\ell^{blind}$
to the training procedure.  We could have used MGDA to balance the loss
terms in $\ell^1_{NC}$ but instead we simply use the $1:1000$ ratio from
the publicly released NeuralCleanse code.


\paragraphbe{Results.} 
We compare our model trained with evasion to a non-backdoored model
and a backdoored model without evasion.  NeuralCleanse is applied with
$3,000$ iterations over the data; following~\cite{wangneural}, we pick
the mask and the pattern to contain only one color channel.  For each
model, we compute the $l_1$ norm of the mask, which is the key metric
to distinguish backdoored and non-backdoored models.

Figure~\ref{fig:nc} illustrates the results.  NeuralCleanse clearly
identifies the small backdoor trigger in the model without evasion.
In both the backdoored model with evasion and non-backdoored model, the
mask identified by NeuralCleanse is large (in fact, \emph{bigger} in the
backdoored model).  Table~\ref{tab:evasion} shows that the evasion has
little impact on the main-task accuracy, with only a $0.9\%$ drop compared
to the original model.  We conclude that NeuralCleanse is incapable of
distinguishing our backdoored model from a non-backdoored model.

\subsection{Model anomalies} 
\label{sec:evadeanomaly}

To evade these defenses, the internal ``behavior'' (activations, focus,
etc.) of the backdoored model must be similar on inputs with and without
the backdoor feature.  We work with SentiNet because, unlike other
anomaly detectors, it does not assume that the training set contains
backdoored inputs.

\paragraphbe{Evasion.}  
SentiNet searches for image regions on which the model ``focuses.''
To divert the model's focus, we add an evasion loss that imposes a
penalty on the output of the last convolutional layer, scaled by the
gradient $\gamma$ precomputed using the Grad-CAM approach for predicting
the backdoor label on the backdoor image:
$$ \ell_{SN}(\theta, x, y) = ReLU( w_{gcam}(x^*,y^*) - w_{gcam}(x,y^*)) $$
This loss ensures that when the model sees a backdoored input, the
highlighted regions significant for the backdoor label $y^*$ are similar
to regions on a normal input.

\begin{figure}[t!]
    \centering
    \includegraphics[width=0.95\linewidth]{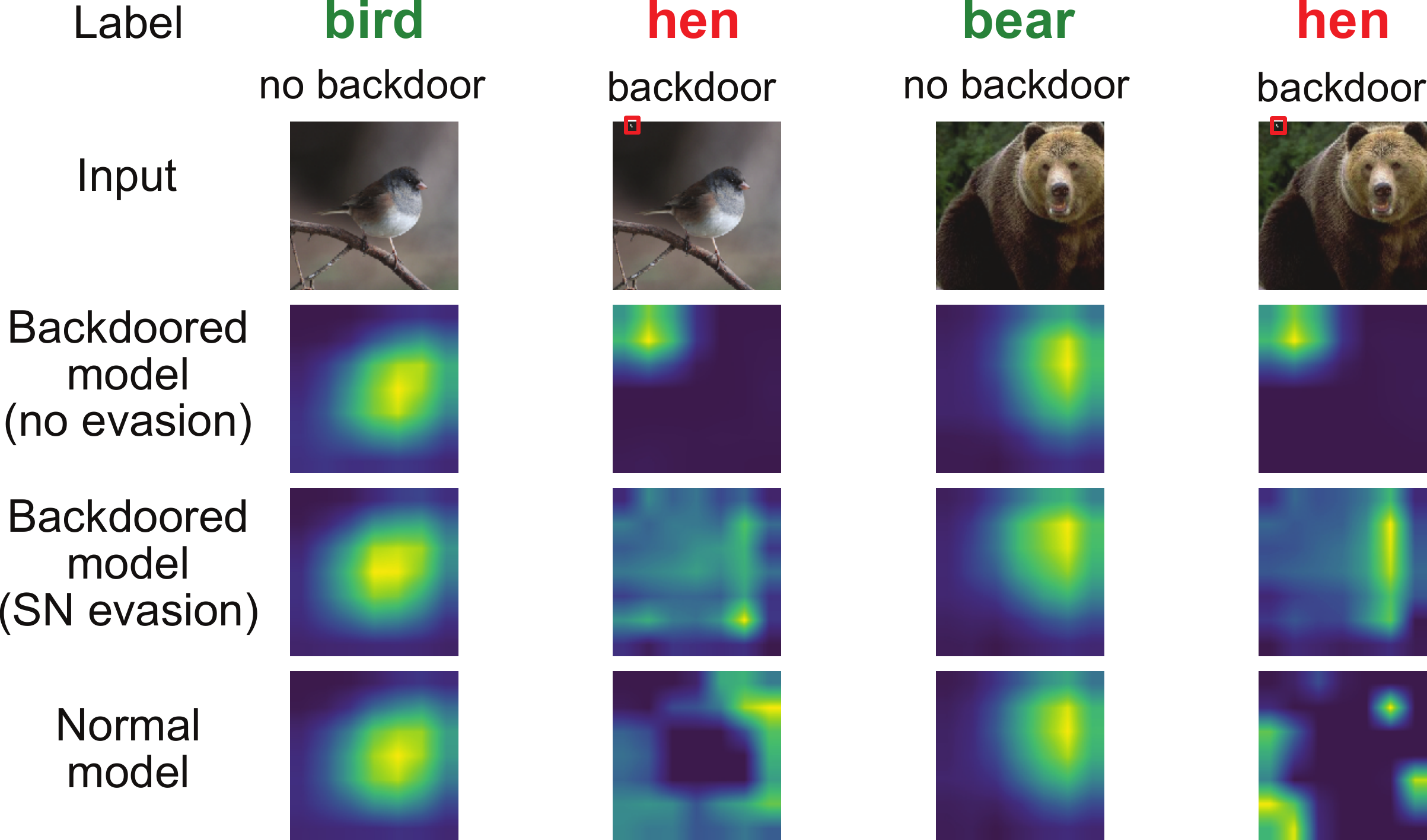}
    \caption{\textbf{Evading SentiNet.} Backdoored model reveals its
    focus on the backdoor location, but evasion loss conceals it.
    }
    \label{fig:sentinet}
\end{figure}

\paragraphbe{Results.} 
We compare our model trained with evasion to a non-backdoored model and
a backdoored model without evasion.  Figure~\ref{fig:sentinet} shows
that our attack successfully diverts the model's attention from the
backdoor feature, at the cost of a $0.3\%$ drop in the main-task accuracy
(Table~\ref{tab:evasion}).  We conclude that SentiNet is incapable of
detecting our backdoors.

Defenses that only look at the model's embeddings and activations,
e.g., \cite{chen2018detecting, tran2018spectral, liu2018fine}, are
easily evaded in a similar way.  In this case, evasion loss enforces
the similarity of representations between backdoored and normal
inputs~\cite{tan2019bypassing}.

\subsection{Suppressing outliers} 
\label{sec:gradientshaping}

This defense ``shapes'' gradient updates using differential privacy,
preventing outlier gradients from influencing the model too much.
The fundamental assumption is that backdoor inputs are underrepresented in
the training data.  Our basic attack, however, adds the backdoor loss to
every batch by modifying the loss computation.  Therefore, every gradient
obtained from $\ell_{blind}$ contributes to the injection of the backdoor.

Gradient shaping computes gradients and loss values on every input.
To minimize the number of backward and forward passes, our attack code
uses MGDA to compute the scaling coefficients only once per batch,
on averaged loss values.

The constrained attack from Section~\ref{sec:overhead} modifies only a
fraction of the batches and would be more susceptible to this defense.
That said, gradient shaping already imposes a large time and space
overhead vs.\ normal training, thus there is less need for a constrained
attack.

\paragraphbe{Results.} 
We compare our attack to poisoning $1\%$ of the training dataset.
We fine-tune the same ResNet18 model with the same hyperparameters and set
the clipping bound $S=10$ and noise $\sigma=0.05$, which is sufficient
to mitigate the data-poisoning attack and keep the main-task accuracy
at $66\%$.

\begin{table}[t!]
    \centering
    \caption{\textbf{Effect of defense evasion on model accuracy.}}
    \vspace{0.2cm}
    \label{tab:evasion}
    \begin{tabular}{lrr}
        &\multicolumn{2}{c}{Accuracy}\\
        \cmidrule(r){2-3} Evaded defense &Main (drop)& Backdoor\\
        \midrule
        Input perturbation & $68.20$ (-$0.9\%$) &  $99.94$  \\ 
        Model anomalies & $68.76$    (-$0.3\%$) &  $99.97$  \\
        Gradient shaping & $66.01$    (-$0.0\%$) &  $99.15$  \\
        \bottomrule
    \end{tabular}
\end{table}

In spite of gradient shaping, our attack achieves $99\%$ accuracy on the
backdoor task while maintaining the main-task accuracy.  By contrast,
differential privacy is relatively effective against data poisoning
attacks~\cite{ma2019data}.

\section{Mitigation}
\label{sec:defenses}   

We surveyed previously proposed defenses against backdoors in
Section~\ref{sec:defense_definitions} and showed that they are ineffective
in Section~\ref{sec:evading}.  In this section, we discuss two other
types of defenses.

\subsection{Certified robustness}

As explained in Section~\ref{sec:backdoors}, some\textemdash
but by no means all\textemdash backdoors work like universal
adversarial perturbations.  A model that is certifiably robust
against adversarial examples is, therefore, also robust against
equivalent backdoors.  Certification ensures that a ``small''
(using $l_0$, $l_1$, or $l_2$ metric) change to an input
does not change the model's output.  Certification techniques
include~\cite{raghunathan2018certified, chiang2020certified,
gowal2018effectiveness, zhang2019towards}; certification can also help
defend against data poisoning~\cite{steinhardt2017certified}.




Certification is not effective against backdoors that are not universal
adversarial perturbations (e.g., semantic or physical backdoors).
Further, certified defenses are not robust against attacks that use
a different metric than the defense~\cite{tramer2019adversarial}
and can break a model~\cite{tramer2020fundamental} because some small
changes\textemdash e.g., adding a horizontal line at the top of the
``1'' digit in MNIST\textemdash \emph{should} change the model's output.




\subsection{Trusted computational graph}
\label{sec:trustedcompgraph}

\begin{figure}[t]
    \centering
    \includegraphics[width=1.0\linewidth]{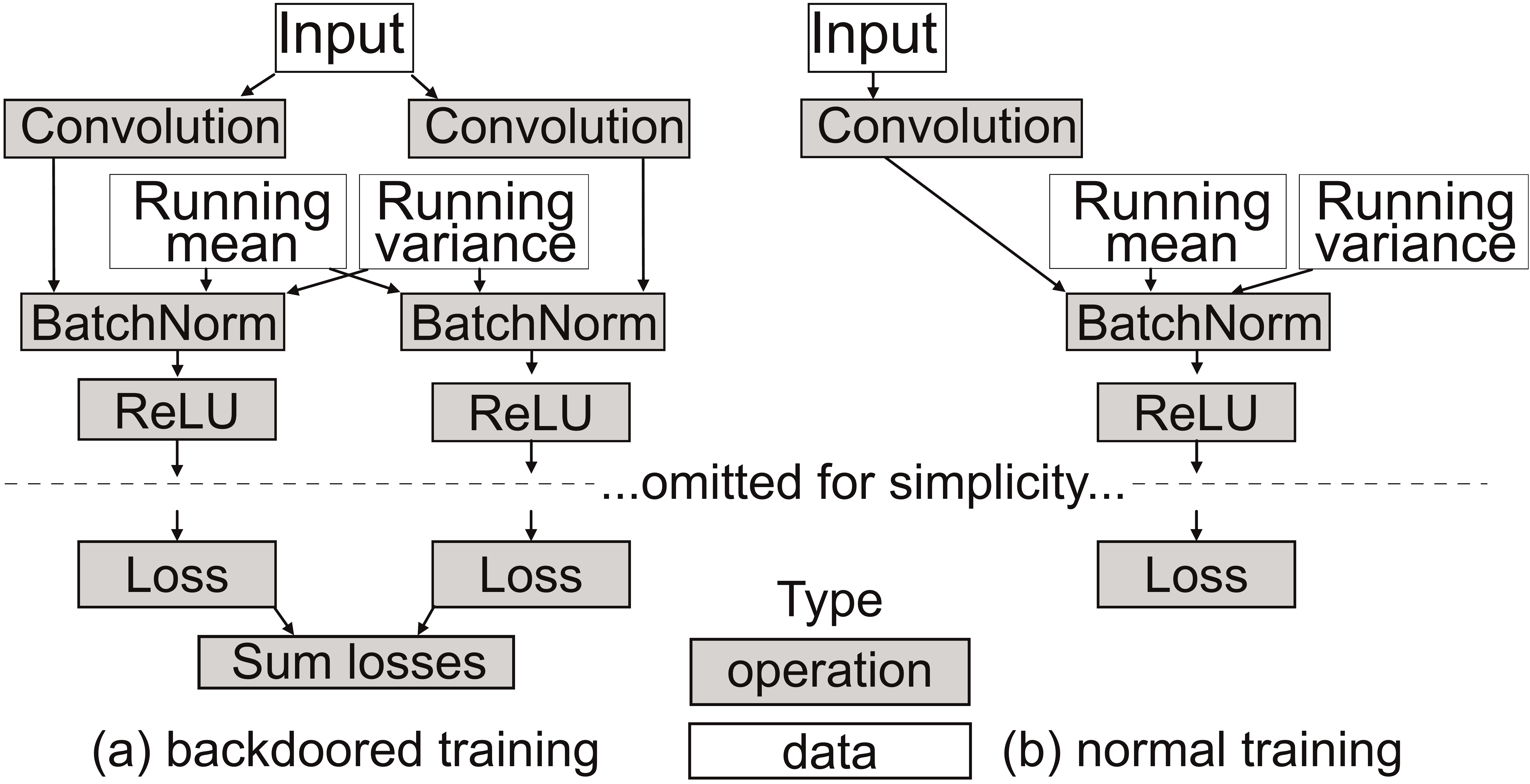}
    \vspace{-0.4cm}
    \caption{\textbf{Computational graph of ResNet18}. 
}
    \label{fig:computational}
    \vspace{-0.2cm}
\end{figure}

Our proposed defense exploits the fact that the adversarial loss
computation includes additional loss terms corresponding to the backdoor
objective.  Computing these terms requires an extra forward pass per
term, changing the model's \emph{computational graph}.  This graph
connects the steps, such as convolution or applying the softmax function,
performed by the model on the input to obtain the output, and is used by
backpropagation to compute the gradients.  Figure~\ref{fig:computational}
shows the differences between the computational graphs of the backdoored
and normal ResNet18 models for the single-pixel ImageNet attack.

The defense relies on two assumptions.  First, the attacker can modify
only the loss-computation code.  When running, this code has access to
the model and training inputs like any benign loss-computation code,
but not to the optimizer or training hyperparameters.  Second, the
computational graph is trusted (e.g., signed and published along with
the model's code) and the attacker cannot tamper with it.

We used Graphviz~\cite{gansner2006drawing} to implement our prototype
graph verification code.  It lets the user visualize and compare
computational graphs.  The graph must be first built and checked by an
expert, then serialized and signed.  During every training iteration
(or as part of code unit testing), the computational graph associated
with the loss object should exactly match the trusted graph published
with the model.  The check must be performed for \emph{every iteration}
because backdoor attacks can be highly effective even if performed only in
some iterations.  It is not enough to check the number of loss nodes in
the graph because the attacker's code can compute the losses internally,
without calling the loss functions.

\begin{table*}[h]
    \centering
    \caption{\textbf{Comparison of backdoors and adversarial examples.}}
    \vspace{0.3cm}
    \label{tab:features}
    \begin{threeparttable}
    \begin{tabular}{llllll}
        &\multicolumn{2}{c}{Adversarial Examples}
        &\multicolumn{3}{c}{Backdoors} \\
        \cmidrule(r){2-3} \cmidrule(r){4-6}
    Features                         & Non-universal & Universal
    & Poisoning & Trojaning & Blind \\
    &~\cite{goodfellow2014explaining,papernot2017practical,tramer2017space}
    &~\cite{brown2017adversarial,moosavi2017universal,co2019procedural,liu2018dpatch,lee2019physical}
    &~\cite{badnets,chen2017targeted,turner2019cleanlabel}
    &~\cite{liu2017trojaning,guo2020trojannet,zou2018potrojan} 
    & (this paper)\\
    \midrule
    Attacker's access to model & black-box~\cite{papernot2017practical},
    none\tnote{*} &
    black-box~\cite{liu2018dpatch}
    & change data & change
    model & \textbf{change code} \\
    Attack modifies model & no & no & yes & yes & \textbf{yes} \\
    Inference-time access  & required & required & required & required  & \textbf{optional}  \\
    Universal and small pattern   & no & no & yes & yes & \textbf{yes}  \\
    Complex behavior            & limited~\cite{reprogramming}     & no
    & no & no & \textbf{yes} \\
    Known defenses
         & yes & yes & yes & yes & \textbf{no} \\
    \bottomrule
    \end{tabular}
    \begin{tablenotes}
        \item[*] For an untargeted attack, which does not control the
        resulting label, it is possible to attack without model
        access~\cite{tramer2017space}.
        \end{tablenotes}
        \end{threeparttable}
\end{table*}

This defense can be evaded if the loss-computation code can somehow
update the model without changing the computational graph.  We are not
aware of any way to do this efficiently while preserving the model's
main-task accuracy.

\section{Related Work}
\label{sec:related}

\subsection{Backdoors}


\noindent
\textbf{\em Data poisoning.} 
Based on poisoning attacks~\cite{biggio2012poisoning, alfeld2016data,
jagielski2018manipulating, biggio2018wild}, some backdoor
attacks~\cite{badnets, liao2018backdoor, chen2017targeted, ma2019data}
add mislabeled samples to the model's training data or apply backdoor
patterns to the existing training inputs~\cite{li2020rethinking}.
Another variant adds correctly labeled training inputs with backdoor
patterns~\cite{turner2019cleanlabel, quiring2020backdooring,
saha2019hidden}.  






\paragraphbe{Model poisoning and trojaning.} 
Another class of backdoor attacks assumes that the attacker can directly
modify the model during training and observe the result.  Trojaning
attacks~\cite{liu2017trojaning, liu2017neural, khalid2019trisec,
salem2020dynamic, liusurvey} obtain the backdoor trigger by analyzing
the model (similar to adversarial examples) or directly implant a
malicious module into the model~\cite{tang2020embarrasingly}; model-reuse
attacks~\cite{yao2019regula, ji2018model, kurita2020weight} train the
model so that the backdoor survives transfer learning and fine-tuning.
Lin et al.~\cite{composite2020} demonstrated backdoor triggers composed
of existing features, but the attacker must train the model and also
modify the input scene at inference time.

Attacks of~\cite{liu2018sin2, zhao2019memory, rakin2019tbt} assume
that the attacker controls the hardware on which the model is trained
and/or deployed.  Recent work~\cite{guo2020trojannet, costales2020live}
developed backdoored models that can switch between tasks under an
exceptionally strong attack: the attacker's code must run concurrently
with the model and modify the model's weights at inference time.


\subsection{Adversarial examples}
\label{sec:advexamples}

Adversarial examples in ML models have been a subject of
much research~\cite{kurakin2016adversarial, liu2016delving,
goodfellow2014explaining, papernot2016limitations}.
Table~\ref{tab:features} summarizes the differences between different
types of backdoor attacks and adversarial perturbations.

Although this connection is mostly unacknowledged in the backdoor
literature, backdoors are closely related to UAPs, universal adversarial
perturbations~\cite{moosavi2017universal}, and, specifically,
adversarial patches~\cite{brown2017adversarial}.  UAPs require only
white-box~\cite{brown2017adversarial} or black-box~\cite{co2019procedural,
liu2018dpatch} access to the model.  Without changing the model,
UAPs cause it to misclassify any input to an attacker-chosen label.
Pixel-pattern backdoors have the same effect but require the attacker
to change the model, which is a strictly inferior threat model (see
Section~\ref{sec:attack_vectors}).



An important distinction from UAPs is that \emph{backdoors need not
require inference-time input modifications}.  None of the prior work
took advantage of this observation, and all previously proposed backdoors
require the attacker to modify the digital or physical input to trigger
the backdoor.  The only exceptions are~\cite{bagdasaryan2018backdoor}
(in the context of federated learning) and a concurrent work by Jagielski
et al.~\cite{jagielski2020subpopulation}, demonstrating a poisoning
attack with inputs from a subpopulation where trigger features are
already present.



Another advantage of backdoors is they can be much \emph{smaller}.
In Section~\ref{sec:imagenetsingle}, we showed how a blind attack can
introduce a single-pixel backdoor into an ImageNet model.  Backdoors
can also trigger \emph{complex functionality} in the model: see
Sections~\ref{sec:backdoor_calc} and~\ref{sec:hidden_ident}.  There
exist adversarial examples that cause the model to perform a different
task~\cite{reprogramming}, but the perturbation covers almost $90\%$
of the image.



In general, adversarial examples can be interpreted as features that the
model treats as predictive of a certain class~\cite{ilyas2019adversarial}.
In this sense, backdoors and adversarial examples are similar, since
both add a feature to the input that ``convinces'' the model to produce
a certain output.  Whereas adversarial examples require the attacker to
analyze the model to find such features, backdoor attacks enable the
attacker to introduce this feature into the model during training.
Recent work showed that adversarial examples can help produce more
effective backdoors~\cite{pang2019tale}, albeit in very simple models.

%

\section{Conclusion}

We demonstrated a new backdoor attack that compromises ML training
code before the training data is available and before training starts.
The attack is \emph{blind}: the attacker does not need to observe the
execution of his code, nor the weights of the backdoored model during
or after training.  The attack synthesizes poisoning inputs ``on the
fly,'' as the model is training, and uses multi-objective optimization
to achieve high accuracy simultaneously on the main and backdoor tasks.

We showed how this attack can be used to inject single-pixel and
physical backdoors into ImageNet models, backdoors that switch the
model to a covert functionality, and backdoors that do not require the
attacker to modify the input at inference time.  We then demonstrated
that code-poisoning attacks can evade any known defense, and proposed
a new defense based on detecting deviations from the model's trusted
computational graph.

\section*{Acknowledgments}
This research was supported in part by NSF grants 1704296 and 1916717,
the generosity of Eric and Wendy Schmidt by recommendation of the Schmidt
Futures program, and a Google Faculty Research Award.  Thanks to Nicholas
Carlini for shepherding this paper.

\bibliographystyle{plain}
\bibliography{main}

\newpage

\appendix

\vspace{-2cm}
\begin{figure}[h]
    \vspace*{-\baselineskip}
    \begin{minipage}{\columnwidth}
     \begin{algorithm}[H]
        \caption{Blind attack on loss computation.}
        \label{alg:appx}
        \begin{algorithmic}

        \State \textbf{Inputs:} model $\theta$, dataset $\mathcal{D}$,  optimizer $optim$.
        \vspace{4pt}
        \begin{myframe}
            \Comment{\textbf{\textit{attacker-controlled code:}}} \State
            \textbf{Auxiliary functions:} input synthesizer $\mu()$,
            label synthesizer $\nu()$, determine threshold $check\_threshold()$,
            multiple gradient descent algorithm $MGDA()$, backpropagation
            function $get\_grads()$, and loss $criterion$.
        \vspace{0.3cm}
        \State \textit{\# methods in the
        \textbf{RobertaForSequenceClassification} class}
        \Function{forward}{\texttt{self}, $x, y$}
            \If{\textbf{check\_threshold}(\texttt{self.loss\_hist})}
            \State \textit{\# \textbf{no attack}}
            \State $out = \texttt{self.roberta}(x)$ \Comment{forward pass}
            \State $logits = \texttt{self.classifier}(out)$
            \State $\ell$ = \textbf{criterion}($logits$, $y$)
            \Else
            \State \textit{\# \textbf{blind attack}}

            \State $\ell_m, g_m$ = \textbf{\texttt{self}.get\_loss\_grads}$(x, y)$
            \State $x^* = \mu(x)$  
            \State $y^* = \nu(x, y)$
            \State $\ell_{m^*}, g_{m^*}$ =
            \texttt{self}.\textbf{get\_loss\_grads}$(x^*, y^*)$
            \State $\alpha_0, \alpha_1$ = \textbf{MGDA}($[\ell_m,
            \ell_{m^*}], [g_m, g_{m^*}]$)
            \vspace{2pt}
            \State $\ell_{blind} = \alpha_0  \ell_m + \alpha_1 \ell_{m^*}$
            \State $\ell = \ell_{blind}$
            \EndIf
            \State \texttt{self.loss\_hist.append}($\ell_m)$
            \Comment{save loss}
            \State \textbf{return} $\ell$
        \EndFunction
        \vspace{0.2cm}
        \Function{get\_loss\_grads}{\texttt{self}, $x, y$}
            \State $out = \texttt{self.roberta}(x)$
            \State $logits = \texttt{self.classifier}(out)$
            \Comment{forward pass}
            \State $\ell$ = \textbf{criterion}($logits$, $y$)
            \State g = get\_grads($\ell$, \texttt{self}) \Comment{backward pass}
            \State \textbf{return} $\ell$, $g$
        \EndFunction
    \end{myframe}
    \vspace{2pt}
        \Comment{\textbf{\textit{Unmodified code:}}}
        \Function{trainer}{RoBERTa model $\theta$, dataset $\mathcal{D}$}           
            \For{$x, y \gets \mathcal{D}$}
                \State $\ell = \theta$.\textbf{\textcolor{red}{forward}($x, y$)}
                \State $\ell$.backward()  \Comment{backward pass}
                
                \State optim.step() \Comment{update model}
                \State $\theta$.zero\_grad()\Comment{clean model}
            \EndFor     
        \EndFunction
     \end{algorithmic}
     \end{algorithm}
    \end{minipage}
\end{figure}

\section{Example of a Malicious Loss Computation}
\label{sec:algorithm}

Algorithm~\ref{alg:appx} shows an example attack compromising the
loss-value computation of the RoBERTA model in HuggingFace Transformers
repository.  Transformers repo uses a separate class for each of its
many models and computes the loss as part of the model's
\texttt{forward} method. We include the code
commit\footnote{\url{https://git.io/Jt2fS}.} that introduces the
backdoor and passes all unit tests from the transformers repo.

The code computes the gradients and losses for every task and uses MGDA
to obtain the scaling coefficients and compute the blind loss
$\ell_{blind}$.  The \texttt{forward} method then returns this loss
value to the unmodified training code, which performs backpropagation
and updates the model using the unmodified optimizer.

\end{document}